\documentclass[12pt]{article}

\usepackage[textsize=tiny]{todonotes}

\usepackage{dcolumn}
\usepackage{amsbsy}
\usepackage{amsmath}

\usepackage{amssymb}
\usepackage{braket}
\usepackage{color}
\usepackage{fullpage}
\usepackage{graphicx}
\usepackage{caption}
\usepackage{bm}

\usepackage{subfigure}

\usepackage{comment}

\newcommand{\ba}{\bm{a}}
\newcommand{\bb}{\bm{b}}

\newcommand{\bg}{\bm{g}}
\newcommand{\bk}{\bm{k}}
\newcommand{\bmm}{\bm{m}}
\newcommand{\bn}{\bm{n}}

\newcommand{\bu}{\bm{u}}
\newcommand{\bv}{\bm{v}}

\newcommand{\bx}{\bm{x}}

\newcommand{\bH}{\bm{H}}
\newcommand{\bI}{\bm{I}}
\newcommand{\bK}{\bm{K}}
\newcommand{\bM}{\bm{M}}
\newcommand{\bP}{\bm{P}}

\newcommand{\bR}{\bm{R}}

\newcommand{\bT}{\bm{T}}

\newcommand{\bW}{\bm{W}}

\newcommand{\bzero}{\mathbf{0}}

\newcommand{\bkappa}{\boldsymbol{\kappa}}

\begin{document}
\title{Edge waves in plates with resonators: An elastic analogue of the quantum valley Hall effect}
\author{Raj Kumar Pal$^{a,*}$, Massimo Ruzzene$^{a,b}$ \\
{\small $^a$ School of Aerospace Engineering, Georgia Institute of Technology, Atlanta GA 30332}\\
{\small $^b$ School of Mechanical Engineering, Georgia Institute of Technology, Atlanta GA 30332}\\
{\small$^*$Corresponding author. E-mail: raj.pal@aerospace.gatech.edu}}

\date{}

\maketitle

\begin{abstract}
We investigate elastic periodic structures characterized by 
topologically nontrivial bandgaps supporting backscattering suppressed edge waves. These edge waves are topologically protected 
and are obtained by breaking inversion symmetry within the unit cell. Examples for discrete one and two-dimensional lattices elucidate the concept and illustrate parallels with the quantum valley Hall effect. The concept is implemented on an elastic plate
featuring an array of resonators arranged according to a hexagonal topology. The resulting continuous structures 
have non-trivial bandgaps supporting edge waves at the interface between two media having different topological invariants. The topological properties of the considered configurations are predicted by unit cell and finite strip dispersion analyses. Numerical simulations on finite structures demonstrate edge wave propagation for excitation at frequencies belonging to the bulk bandgaps. The considered plate configurations  define a framework for the implementation of topological concepts on continuous elastic structures of potential engineering relevance. 
\end{abstract}

\section{Introduction}
The study of  topologically protected phenomena in materials and metamaterials is an active area of research that draws inspiration from 
quantum systems~\cite{hasan2010colloquium,haldane1988model}. Recent developments include 
classical areas such as acoustics~\cite{yang2015topological}, 
optomechanics~\cite{peano2014topological,peano2016topological}, elastic~\cite{chen2016topological,sussman2015topological}
and photonic systems~\cite{haldane2008possible,lu2014topological}. 
The topological properties of the band structure, that is, electronic 
bands in quantum mechanics or dispersion surfaces in photonic, acoustic and mechanical systems~\cite{rocklin2016mechanical}, can be exploited to achieve unique and exciting properties. The exploration of such properties has motivated the development of classification schemes for the various types of topological phases that are available~\cite{susstrunk2016classification,ryu2010topological}.  
One such property is the existence of edge waves at interfaces or boundaries, and these waves are immune to backscattering in the presence of a broad class of imperfections and impurities, including localized defects and sharp corners. 

Topologically protected wave propagation is supported by systems belonging to two broad categories. The first one 
relies on breaking time reversal symmetry to produce chiral edge modes and it generally requires active components or the application of external fields.  For example, Prodan and Prodan~\cite{prodan2009topological} demonstrated how 
having weak magnetic forces which break time-reversal symmetry 
can induce edge modes in biological systems, while, more recently, 
Wang et al.~\cite{wang2015topological} used rotating gyroscopes as a way to break time reversal symmetry. Other examples involved the rotation of the entire lattice~\cite{kariyado2015manipulation}, rotating disks at each location~\cite{nash2015topological}, 
or springs with time modulated constants leading to non-reciprocal wave motion~\cite{khanikaev2015topologically,swinteck2015bulk}. 
For these $2D$ lattice systems, analogies can be drawn to the quantum Hall effect~\cite{haldane1988model}.  A second category exhibits helical edge modes in analogy with the quantum spin Hall effect~\cite{hasan2010colloquium}.  These systems do not break time reversal symmetry and solely employ passive components. Examples include the configuration investigated by S{\"u}sstrunk and Huber~\cite{susstrunk2015observation},  who experimentally demonstrated topologically protected edge waves in Hofstadter lattices consisting of a combination of springs and levers connected to linear pendulums. 
Khanikaev et al.~\cite{mousavi2015topologically} conducted numerical studies on wave propagation
in a plate having $2$-scale perforations, while Chen and coworkers~\cite{he2015acoustic} 
achieved helical edge modes at the interface of two lattices including two distinct sizes of steel cylindrical inclusions  in water. Similarly, Pal et al.~\cite{pal2016helical} obtained helical edge waves in a bi-layer mechanical lattice consisting of a combination of regular and chiral springs. Other examples include coupled pendula~\cite{salerno2016spin}, electrical 
~\cite{ningyuan2015time} and piezoelectric systems~\cite{mchugh2016topological}.  Most of these configurations consist of discrete systems that, while suitable for describing basic concepts, are not immediately transferable to physical configurations that may lead to practical applications.

This work investigates elastic continuous structures that emulate the quantum valley Hall effect (QVHE) to achieve topologically protected edge modes. The QVHE exploits valley states instead of spin states, with the advantage that each lattice site needs to have only one degree of freedom. This concept provides the opportunity to obtain configurations of reduced geometrical complexity. Valley degrees of freedom arise naturally in systems with time reversal symmetry and have been predicted theoretically in graphene~\cite{rycerz2007valley,xiao2007valley}, where wave-functions at opposite valleys feature opposite polarizations and thus emulate spin orbit interactions. This concept was extended by Ma and Shvets~\cite{ma2016all} to a photonic crystal exhibiting topologically protected valley edge states, while Dong and coworkers~\cite{chen2016valley}  illustrated valley modes in photonic crystals with an hexagonal lattice of inclusions. Recently, this concept has also been extended to acoustic waves propagating in a phononic crystal, where triangular stubs provide the opportunity to break inversion symmetry by varying their orientation with the lattice~\cite{lu2016valley}. 

In continuous structures like plates and shells, wave guiding through edge modes remains an open challenge, one that, if solved, could have important applications for wave isolation, impact mitigation and the transfer of information through elastic waves. Implications could affect a diverse range of fields such as acoustic imaging, SAW devices, noise control and energy harvesting. In recent years, using analogies with discrete systems, bandgaps have been induced in plates through array of resonators. The resonators have been  idealized as spring mass systems in theoretical studies~\cite{xiao2012flexural,torrent2013elastic} or have been physically implemented by cutting holes in periodic arrangements~\cite{andreassen2015directional}. Other studies have considered 
surface mounted stubs of single and multiple materials acting as both Bragg scatterers and internal resonators~\cite{oudich2011experimental}. 

This work considers resonators on elastic plates as an effective strategy to induce topologically protected wave propagation in a continuous elastic system such as a plate. In contrast with numerous studies in the 
past decade considering phononic crystals and periodic media where waveguiding is susceptible to localization and  backscattering at defects and imperfections, the proposed approach leads to backscattering suppressed edge waves. The approach is illustrated on two distinct but related discrete lattices to finally lead the development of a plate with resonators. The outline of the paper is as follows. Following this introduction (Section1), Section~\ref{sec:theory} presents a  description of the 
lattices and the associated theory, drawing parallels with existing work in related areas to demonstrate the generation of edge modes. 
Section~\ref{sec:numerResult} describes the implementation of the concept on elastic plates, for which dispersion studies are first conducted and subsequently verified through numerical simulations. The main results of the work and conclusions are finally summarized in Section~\ref{sec:conc}. 

\section{Background: Discrete lattices}\label{sec:theory}
\subsection{Band inversion and quantum valley Hall effect}
It is well known that a $1D$ periodic lattice with distinct stiffness or mass values within a unit cell is characterized by a frequency bandgap~\cite{hussein2014dynamics}
and that the bands are characterized by a topological invariant~\cite{xiao2014surface}. 
Furthermore, recent studies~\cite{xiao2014surface} have shown that at 
the interface of two lattices with distinct topological invariants, a topologically protected localized edge mode exists. 
In a $2D$ lattice with Dirac points, breaking inversion
symmetry while preserving $C_3$ symmetry can lead to topologically protected edge modes. These modes are  
helical in nature~\cite{xiao2007valley,ma2016all} and are associated with the quantum valley Hall effect. 

We construct lattices which exhibit topologically protected edge modes by changing parity of the springs or by
breaking inversion symmetry within the unit cell. The principle 
is illustrated through $2$ simple examples on discrete lattices. The first example involves a simple spring mass chain where localized mode forms at the interface between two lattices with distinct topological indices. We then present a  $2D$ extension where topologically protected edge waves exist at the interface between lattices with two different material parameters.  Non-trivial topological edge modes are illustrated by calculating the invariants associated with the bands. The examples also show how the effective Hamiltonian is equivalent to the form found in other studies to illustrate the analogy with the quantum valley Hall effect.

\subsection{One Dimensional Lattice}

\begin{figure}
	\centering
	\subfigure[]{
	\includegraphics[width=0.35\linewidth]{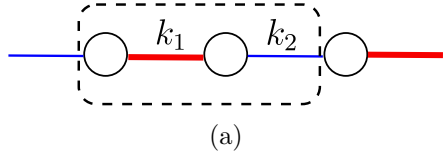}
	\label{fig:1Dscheme}
	}\\
	\subfigure[]{
	\includegraphics[width=0.5\linewidth]{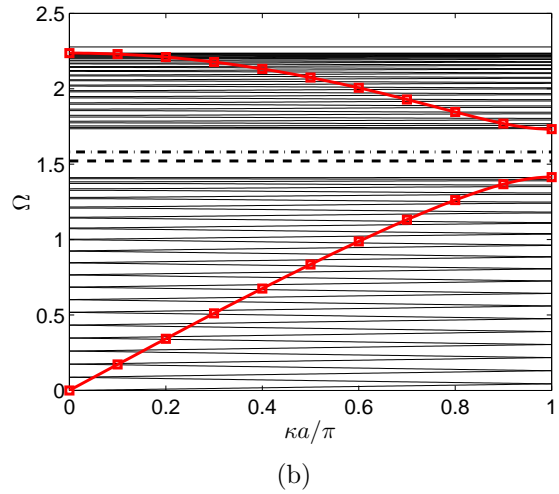}
	\label{fig:1Dlattice}
	}
	\caption{(a) Schematic of the $1D$ lattice with distinct adjacent springs. 
	(b) Dispersion diagram for a single unit cell (red dashed lines) and a finite number of both types of unit cells (black solid lines). 	Two localized modes having frequency in bandgap arise at the interface corresponding to the two interface types. 
	}
	\label{fig:Schematics}
\end{figure}

Consider the simple one dimensional (1D) spring mass chain in Fig.~\ref{fig:1Dscheme}. The masses are all identical while the stiffness of the springs 
alternate between $k_1$ and $k_2$. Let $m$  and $a$ denote the mass and distance between adjacent masses, respectively. 
The governing equations of a lattice unit cell are 
\begin{subequations}
\begin{align}
m \ddot{u}^a_p + k_1\left(u^a_{p} - u^b_{p}\right) + k_2\left(u^a_{p} - u^b_{p-1}\right) &= 0, \\
m \ddot{u}^b_p + k_1\left(u^b_{p} - u^a_{p}\right) + k_2\left(u^b_{p} - u^a_{p+1}\right) &= 0. 
\end{align}\label{eq: 1Dlattice governing equation}
\end{subequations}
Plane wave propagation is investigated by imposing 
a solution of the form $\bu_p(t) = \bu_0 e^{i(\omega t- \kappa a p)}$, where $\kappa$ is the wavenumber and 
$\bu_p = [u_p^a,\;u_p^b]$. Substituting this expression in 
Eqn.~\eqref{eq: 1Dlattice governing equation} leads to an eigenvalue problem in terms of $\omega$ for a given wavenumber $\kappa$: 
$\bH(\kappa)\bu(\kappa)= \omega^2 m\bu(\kappa)$, with
\begin{equation}
\bH = \begin{bmatrix} k_1 + k_2 & -k_1 - k_2 e^{-i\kappa a} \\  -k_1 - k_2 e^{i\kappa a} & k_1 + k_2 \end{bmatrix}. 
\end{equation}

We show the presence of a localized mode at the interface of two lattices whose parity of springs is flipped. 
Two kinds of unit cells are defined based on the relative values of $k_1$ and $k_2$. We denote a unit cell of type $A$ for $k_1 > k_2$, while a 
type $B$ is obtained for $k_1 < k_2$. We first consider the wave propagation in an infinite chain by analyzing the band structure of a single unit. This provides a reference with the behavior of a conceptual bulk material that is represented by the lattice in the considered configuration. Next, extended domains
consisting of $N=10$ contiguous unit cells of type $A$, connected to $N=10$ type $B$ cells  is considered. The extended domain represents the behavior of a lattice having a single interface with $A$ type cells on one side
and $B$ type cells on the other side of the interface. Note that two kinds 
of such interface conditions exist: the first kind has two adjacent $k_1$ springs, while the second kind features two adjacent $k_2$ springs. The dispersion properties of lattices with both interfaces are evaluated and compared with that of the bulk in Fig.~\ref{fig:1Dlattice}, which shows the dispersion diagrams for the a uniform lattice (bulk) (red lines with square markers) along with the dispersion corresponding to the extended domains (black lines). As expected, the uniform lattice is characterized by an optical and an acoustic branch, separated by a bandgap. The dispersion diagrams for the extended domains having $2 N$ unit cells ($N$ cells of type $A$ and $N$ of type $B$) span wavenumbers $\kappa \in[-\pi a/N , \pi a/N]$ and are here folded on the First Brillouin zone for the bulk. In addition, two additional flat bands exist in the bulk bandgap and correspond to two interface modes associated with the two types of interface mentioned above. 
Specifically, the modes associated with the lower dashed line and higher dot-dashed line are localized at interfaces 
having two adjacent low stiffness and adjacent 
high stiffness springs, respectively. There is also an additional localized mode above the optical 
branch, which is localized at an interface having adjacent high stiffness springs. 

\begin{figure}
	\centering
	\subfigure[]{
	\includegraphics[width=0.45\linewidth]{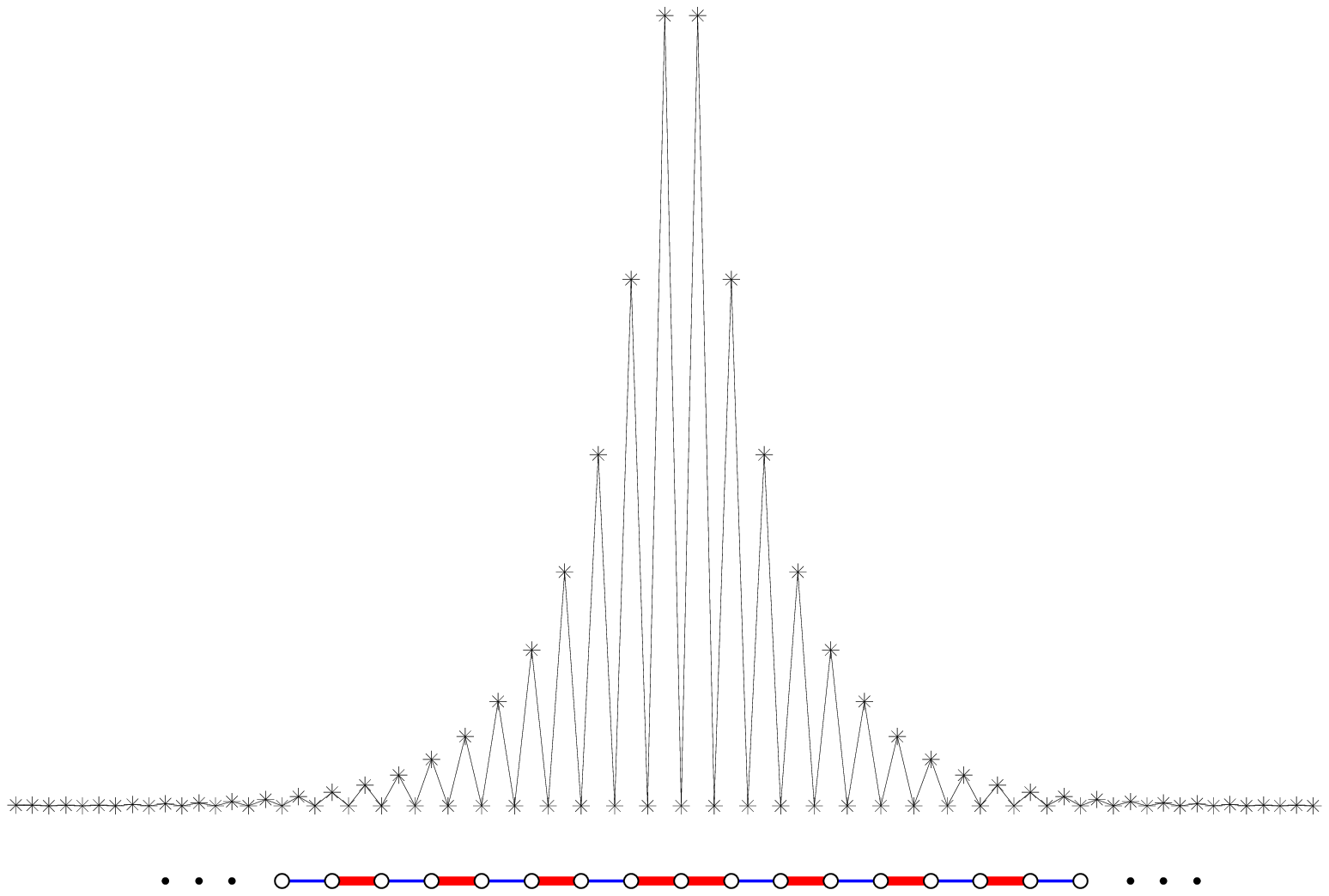}
	\label{fig:HeavyInterface}
	}
	\subfigure[]{
	\includegraphics[width=0.45\linewidth]{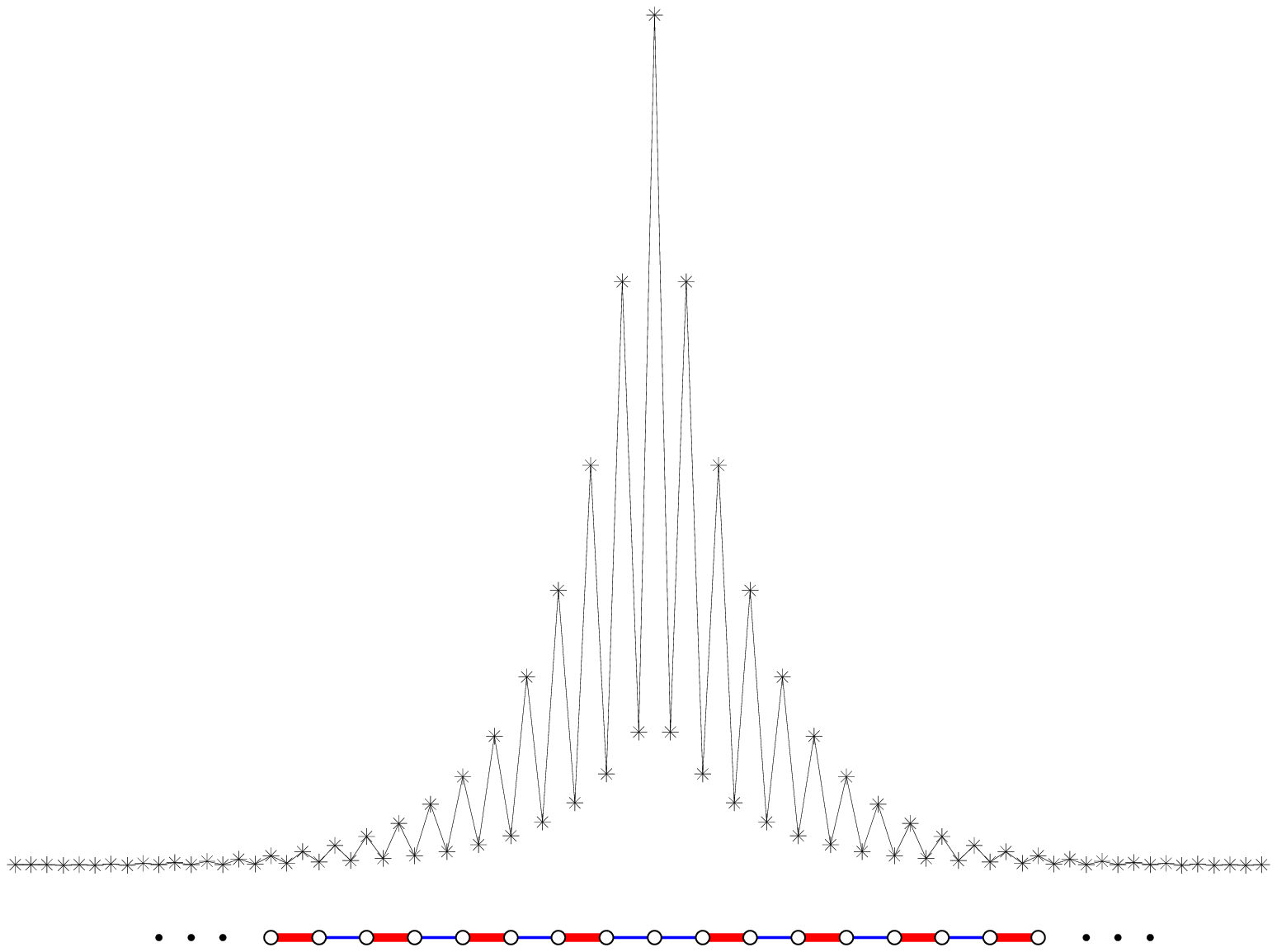}
	\label{fig:LightInterface}
	}
	\caption{Amplitude of localized modes at interface having (a) adjacent high stiffness springs and (b) adjacent low stiffness springs at the interface 
	between two unit cell types.}
	\label{fig:1DinterfaceModes}
\end{figure}

The existence of these interface modes can be predicted by evaluating the topological invariants 
associated with the bulk bands of the lattice. Specifically, we show that
the two cell types, ($A$ and $B$) are characterized by the dissimilar Zak phases~\cite{xiao2014surface}. The Zak phase is a measure of the rotation of the eigenvectors after fixing an appropriate gauge as the wavenumber spans the range $\kappa = -\pi/a:\pi/a$ and 
it is evaluated through the following expression~\cite{zak1989berry}
\begin{equation}
Z = \dfrac{i}{2\pi}\int_{-\pi/a}^{\pi/a} \bu^*(\kappa) \partial_{\kappa} \bu(\kappa) dk. 
\end{equation}
It can be shown that when $k_1 > k_2$ (cell $B$), the Zak phase is $Z=0$ for both acoustic and optical branches, 
while for $k_1 < k_2$ (cell $A$) $Z=1$ for both branches. At an interface between two lattices with distinct Zak phases, a 
localized edge mode forms at a single frequency in the common bandgap of the two lattices~\cite{xiao2014surface,xiao2015geometric}. The difference in Zak phases is further elucidated by considering the problem in 
the basis $\bm{\bar{u}}=[u^s,\; u^w]$, where $u^s = u^a + u^b$ and $u^w = u^a - u^b$ respectively denote a symmetric
and an anti-symmetric mode. In this basis, for $k_1 > k_2$ (cell $B$), the eigenvector at both $\kappa = 0$ and $\kappa=\pi/a$ for the acoustic branch is $\bm{\bar{u}}_{ac} = [1, \; 0]$, while for the optical branch, the 
eigenvector is $\bm{\bar{u}}_{op} =[0, \;1]$. Thus the acoustic and optical branches are associated with symmetric and antisymmetric 
modes,  respectively, at both $\kappa=0$ and $\pi$. 
The modes do not flip in the range $\kappa = -\pi/a:\pi/a$ in both the bands, which corresponds to a Zak phase $Z=0$.
Next, we consider the eigenvectors associated with a lattice having $k_1 < k_2$ (cell $A$). The eigenvector at $\kappa=0$ and 
$\kappa=\pi$ for the acoustic branch are now $\bm{\bar{u}}_{ac} = [1, \; 0]$ and $\bm{\bar{u}}_{ac} = [0, \; 1]$ respectively, while for the optical branch they are  $\bm{\bar{u}}_{op} = [0, \; 1]$ at $\kappa = 0$ and $\bm{\bar{u}}_{op} = [1, \; 0]$ at $\kappa=\pi/a$. 
In contrast with the lattice of type $B$, the eigenvectors interchange between the acoustic and optical branch, which 
corresponds to a Zak phase $Z=1$ for both the bands. 
Thus, although the frequencies in the bulk band structure are identical for the $k_1> k_2$ and the $k_1 < k_2$ lattices, 
the band topology is distinct as quantified by the different Zak phase values. 
Such a difference generates a localized mode at an interface between the two lattice types~\cite{xiao2014surface}. 
The frequency of this mode depends on the type of interface, i.e. if it is of the $k_1$-$k_1$ or the $k_2$-$k_2$ type.

To gain further insight, we examine the displacement field associated with these localized modes. These localized modes are eigenvectors 
corresponding to the bandgap frequencies and are obtained
from the dispersion analysis of the extended unit cell discussed above using spring stiffness $k_1=1$ and $k_2 = 1.5$. Figure~\ref{fig:1DinterfaceModes} 
displays the displacement amplitudes of the masses for 
localized modes which arise at the two kinds of interface along with a zoomed-in view of the interface region below the amplitude plot. 
Note that the eigenvectors of these localized modes are independent of the wavenumber and they are representative of the 
behavior at the interface between two semi-infinite lattices. 
Figure~\ref{fig:1DinterfaceModes} displays the schematics of the chain having two kinds of interface:  $k_1$-$k_1$ and $k_2$-$k_2$ interface. 
The displacement is localized at one of the masses and it decays rapidly away from the interface in both cases. 
Note, however, that the two modes are different and 
the displacement amplitude is higher on different masses. Thus we remark here that localized modes 
always exist between the two lattice types irrespective of the interface type. 
Though these localized modes are similar to defect modes, note that they cannot be removed by varying the properties at the 
interface between the two lattices. The frequencies of the localized modes may change, however, they cannot be moved into the bulk bands.  
This behavior is in contrast with the localized mode which arises above the optical branch as shown in the dispersion diagram 
of an extended unit cell in Fig.~\ref{fig:1Dlattice}. This mode arises at the interface only if 
there are two adjacent heavy springs and is a trivial defect mode. Indeed, by varying the interface type, this mode can be moved into the 
bulk bands and hence it does not arise when the interface has two low stiffness springs.

\subsection{Two dimensional Discrete Lattice}
We now extend the study to a two dimensional (2D) discrete lattice, whereby topologically protected edge modes in a nontrivial bandgap
are obtained by breaking inversion symmetry within the unit cell. The considered hexagonal lattice  consists of point masses at nodes connected by linear springs (Fig.~\ref{fig:2Dscheme}). Each unit cell contains two different masses, respectively equal to $m^a = (1 + \beta) m $ and $m^b = (1-\beta) m$, so that inversion symmetry of the lattice is broken when $\beta \ne 0$, while $C_3$ symmetry  (rotation by $2\pi/3$) is always preserved. Each mass has one degree of freedom corresponding to its out-of-plane motion, while the springs provide a force proportional to the relative motion of connected masses through a constant $k$.
The governing equations for the masses in unit cell $p,q$ are 

\begin{subequations}
\begin{align}
m^a \ddot{u}^a_{p,q} +  k \left( 3 u^a_{p,q} - u^b_{p,q}- u_{p,q-1}^b - u_{p-1,q}^b \right) &= 0, \\ 
m^b \ddot{u}^b_{p,q} +  k \left( 3 u^b_{p,q} - u^a_{p,q}- u_{p,q+1}^a - u_{p+1,q}^a \right) &= 0. 
\end{align}
\end{subequations}

\subsubsection{Dispersion Analysis}
We proceed to seek for plane harmonic waves in the form  $\bm u_{p,q} = \bm{u}_0 e^{i(\omega t + \bkappa \cdot \bm r_{p,q})}$, where $\bm r_{p,q}= p \bm{a}_1 +q \bm{a}_2$ defines the position of the cell $p,q$ in terms of the lattice vectors $\bm{a}_1 = a[1\;0], \bm a_2 =a [\cos(\pi/3) , \; \sin(\pi/3)]$ (where $a=1$ for simplicity), while $\bkappa = \kappa_1 \bg_1 + \kappa_2 \bg_2$ is the wave vector  expressed in the basis of the reciprocal lattice vectors $\bg_1,\bg_2$. Substituting this expression into the governing equations leads to the 
following eigenvalue problem to be solved in terms of  frequency for an assigned wave vector $\bkappa$
\begin{equation}
\Omega^2 \begin{bmatrix} 1+\beta & 0 \\ 0 & 1-\beta \end{bmatrix}
\begin{bmatrix} u^a \\ u^b \end{bmatrix} =  \begin{bmatrix} 3 & -1-e^{-i\bkappa\cdot\ba_1}-e^{-i\bkappa\cdot \ba_2} \\
 -1-e^{i\bkappa\cdot\ba_1}-e^{i\bkappa\cdot \ba_2}  & 3 \end{bmatrix} \begin{bmatrix} u^a \\ u^b \end{bmatrix}. \label{eqn:Eig0}
\end{equation}
Note that a non-dimensional frequency $\Omega^2=\omega^2 m/k$ is introduced for convenience.

To show equivalence with lattices which exhibit edge modes in in quantum, photonic or 
acoustic systems~\cite{xiao2007valley,chen2016valley,ma2016all,lu2016valley}, 
a change of variables to $\bv = \bP \bu$, where $\bu =[u^a, ; u^b]$, is considered, with
\begin{equation}
\bP = \begin{bmatrix} \sqrt{1+\beta} & 0 \\ 0 & \sqrt{1-\beta} \end{bmatrix},
\end{equation}
which effectively corresponds to a stretching of coordinates. Premultiplying both side of the eigenvalue 
problem in Eqn.~\eqref{eqn:Eig0} by $\bP^{-1}$gives
\begin{equation*}
\bH(\bkappa) \bv(\bkappa)= \bP^{-1} \bK(\bkappa) \bP^{-1}\bv(\bkappa) = \Omega^2 \bv(\bkappa).
\end{equation*} 
The matrix $\bH(\bkappa) $ can be written as 
\begin{align}
\bH(\bk) 
&= \begin{bmatrix} 3/1+\beta &  d(\bk)^*/\sqrt{1-\beta^2} \\ d(\bk)/\sqrt{1-\beta^2} & 3/1-\beta   \end{bmatrix} \nonumber\\
&= \left(\dfrac{6}{1-\beta^2}\right)\begin{bmatrix}1 & 0 \\ 0 & 1 \end{bmatrix} +
		\dfrac{1}{\sqrt{ 1-\beta^2 }}\begin{bmatrix}0 & d(\bk)^* \\ d(\bk) & 0 \end{bmatrix} -
		\left(\dfrac{6\beta}{1-\beta^2}\right) \begin{bmatrix}1 & 0 \\ 0 & -1 \end{bmatrix}, 
\end{align}
where $d(\bk) = -1 - e^{i\bkappa\cdot\ba_1} - e^{i\bkappa\cdot \ba_2}$.

The first term in the above expression is a constant times the identity matrix. Its  sole effect is to translate the dispersion bands upward or downward, without  affecting their topology. The second term is similar to the effective mass Hamiltonian of 
graphene~\cite{kane2005quantum} and leads to a Dirac cone 
in the absence of additional interaction terms. Finally, the last term is the result of breaking the inversion symmetry and vanishes when the masses are equal ($\beta=0$). 

\begin{figure}
	\centering
	\subfigure[]{
	\includegraphics[width=0.25\linewidth]{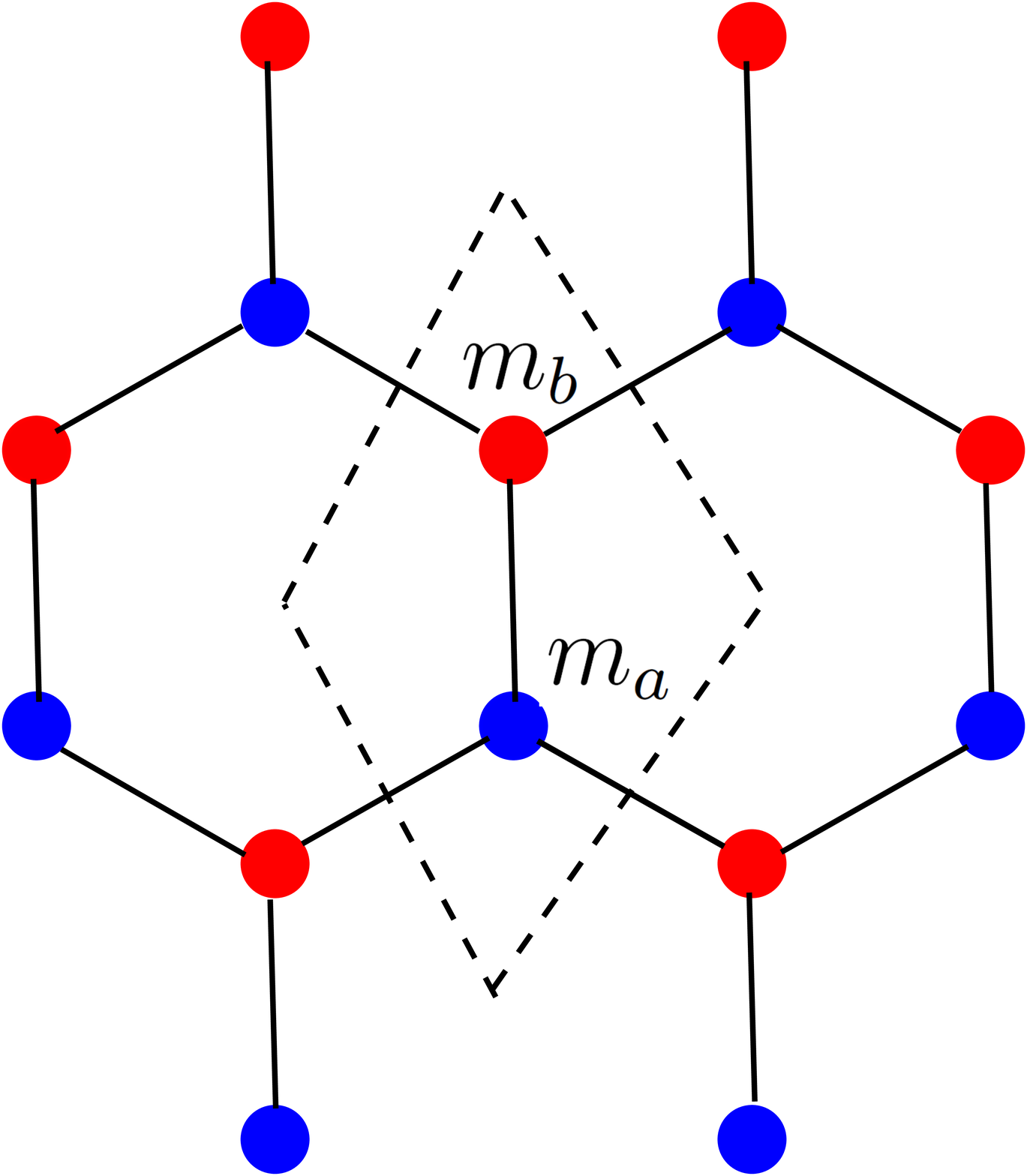}
	\label{fig:2Dscheme}
	}\\
	\subfigure[]{
	\includegraphics[width=0.6\linewidth]{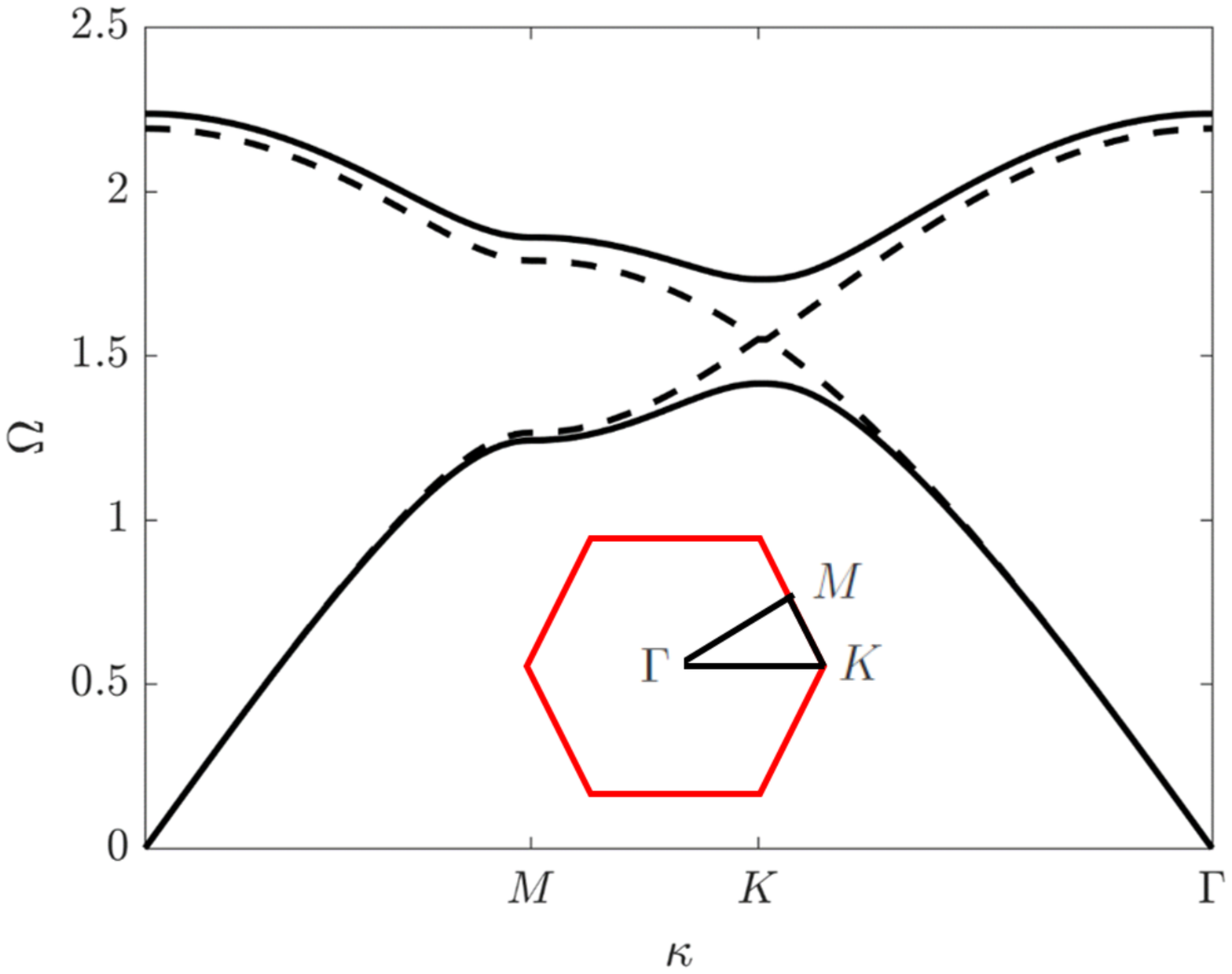}
	\label{fig:2Ddiscrete_IBZ}
	}
	\caption{(a) Schematic of a $2D$ hexagonal lattice having distinct masses in a unit cell, 
	resulting in broken inversion symmetry, but preserved $C_3$ symmetry. 
	(b) Dispersion diagrams along the IBZ for both the unit cell with equal masses  ($\beta=0$) (dashed lines) and the unit cell having dissimilar masses (solid lines) ($\beta=-0.2$). A bandgap opens due to broken inversion symmetry in the latter case. }
\end{figure}

Figure~\ref{fig:2Ddiscrete_IBZ} displays the dispersion diagram for two types of unit cells 
along the corners of the irreducible Brillouin zone (IBZ, sketched as in inset in the figure). 
The dispersion of a unit cell with identical masses $m^a = m^b = 1$ $(\beta=0)$ is shown by dashed curves. A Dirac cone is observed at the $K$ point and features  $6$-fold symmetry. The solid curves show the 
dispersion diagram for a lattice with broken inversion symmetry ($\beta=-0.2$, $m^a = 0.8$  and $m^b=1.2$) for which a bandgap opens at the $K$ point. The dispersion is characterized by broken inversion symmetry, and preserved $C_3$ symmetry. The case obtained with values of the interchanges, i.e. ($\beta=0.2$, $m^a = 1.2$  and $m^b=0.8$) would appear identical, although a band inversion would occur with eigenvectors associated with the corresponding frequencies  flipped.


\begin{figure}
	\centering
	\includegraphics[width=0.5\linewidth]{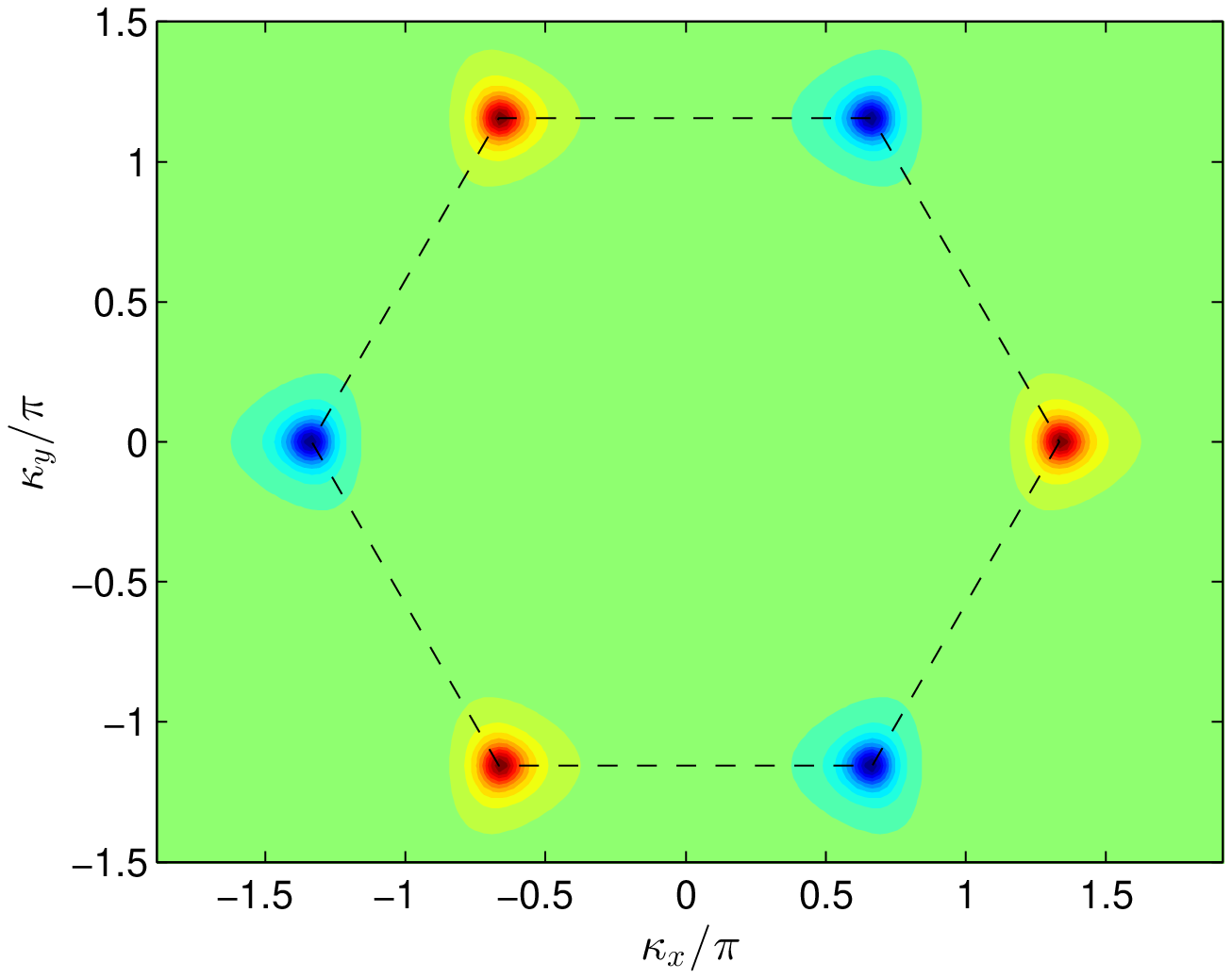}
	\caption{Berry curvature over the first Brillouin zone is localized at the $K$ and $K'$ points. It has opposite signs at these points. }
	\label{Berry2Ddiscrete}
\end{figure}
 
To illustrate the analogy with systems exhibiting quantum valley Hall effect, the above Hamiltonian is expressed in the basis of an extended vector, combining the displacements at the $K$ and $K'$ valley points. These points have Dirac 
cones in the presence of inversion symmetry ($\beta=0$). We consider an extended state vector $\bm{\psi} = [U_K,U_{K'}]$ combining both valleys. Let $\tau_i$ and $\sigma_i$ be Pauli matrices matrices associated with the valley and the unit cell degrees of freedom. The effective Hamiltonian near the Dirac points in this extended basis can then be expressed as 
\begin{equation}
H_0(\delta \bk) = \dfrac{1}{\sqrt{1-\beta^2}} \left( \delta k_x \tau_z \sigma_x + \delta k_y \tau_0 \sigma_y \right) 
+ \left(\dfrac{6}{1-\beta^2} \right) \tau_0 \sigma_0 -
\left(\dfrac{6\beta}{1-\beta^2 } \right) \tau_0 \sigma_z. 
\end{equation}

Alternately, the nontrivial nature of the bands given by the solution of the eigenvalue problem in Eqn.~\eqref{eqn:Eig0} 
can be characterized by computing the associated topological invariants. The relevant topological invariant in this case is the valley Chern number~\cite{xiao2007valley,ma2016all}, which is obtained by integrating 
the Berry phase over half the Brillouin zone. 
We express the eigenvalue relation in Eqn.~\eqref{eqn:Eig0} as $\Omega_m^2\bM \bmm = \bK \bmm$ with $\Omega_m$ being the frequency 
associated with the eigenvector $\bmm$.  
Note that the eigenvectors are normalized to satisfy $\bI = \ket{\bn}\bra{\bM|\bn}$,  where
the bracket notation $\braket{\ba|\bb} = \sum_p a_p^* b_p$ denotes the inner product of the vectors $\ba, \bb$. 
The Berry curvature of a band at wave vector $\bkappa$ having eigenvector $\bmm$ 
is given by $B = i\braket{d\bmm|\bM|d\bmm}$, where $d$ is the exterior derivative operator. 
Differentiating the above eigenvalue relation 
with respect to $\kappa_s$, with $s= \{x,y\}$, and premultiplying by the  eigenvector $\bn$ leads to the following identity:
\begin{equation}
\Braket{\bn |\bM| \dfrac{\partial \bmm}{\partial \kappa_s}} = 
					\dfrac{\Braket{\bn | \partial \bK/\partial \kappa_s | \bmm}}{ \Omega^2_m - \Omega^2_n}.
\end{equation}
We compute the Berry curvature at $\bkappa$  by 
considering an equivalent expression, given in component form as
\begin{align}
B(\bkappa) &= i \braket{d\bmm|\bM|d\bmm} = i \Braket{ \dfrac{\partial \bmm}{\partial \kappa_x} |\bM| \dfrac{\partial \bmm}{\partial \kappa_y} } - c.c 
						= i \sum_{n=1, n\ne m}^N	\Braket{ \dfrac{\partial \bmm}{\partial \kappa_x} | \bM|\bn } 
								\Braket{\bn |\bM| \dfrac{\partial \bmm }{\partial \kappa_y}} - c.c. \nonumber\\  
						&= i\sum_{n=1,n \ne m}^N \dfrac{ \Braket{ \bmm | \dfrac{\partial \bK}{\partial \kappa_x} | \bn } 
						\Braket{ \bn | \dfrac{\partial \bK}{\partial \kappa_y} | \bmm } - c.c. }
						{ \left( \Omega_m^2 - \Omega_n^2 \right)^2 } . \label{eqn:BerryCurv} 
\end{align}
where $c.c.$ denotes the complex conjugate and $N$ denotes the number of eigenmodes. 
Note that the summation in the last step reduces to a single term as we only have $2$ bands for the considered discrete hexagonal lattice. 

Figure~\ref{Berry2Ddiscrete} 
displays the Berry curvature over the entire Brillouin zone. It is localized at the $K$ and $K'$ points, with opposite signs at those points. Furthermore, since the system is time reversal invariant, the total Berry curvature over the whole band is zero. 
The valley Chern number is then computed by integrating the Berry curvature over a small region near the $K,K'$ point as:
\[C_{\nu} = (1/2\pi)\int_{\nu} B(\bkappa)d\bkappa\]
where $\nu = K,K'$ denotes the valley type. Evaluation of the Chern number reveals that for $\beta>0$, i.e. $m^a > m^b$, the lower band is characterized by 
$C_\nu = (-)1/2$ at $K$ ($K'$)  valleys, while opposite signs are found for the upper band. In contrast, the values are reversed, i.e. $C_\nu = (-)1/2$ at $K'$ ($K$) valleys for the lattice with $\beta<0$, i.e. $m^a < m^b$. This demonstrates how 
breaking inversion symmetry by varying the relative masses of the two resonators provides distinct valley Chern numbers for the bands. As discussed in numerous works on quantum and photonic systems, at the interface between two lattices with distinct valley Chern numbers, bulk boundary correspondence guarantees the presence of topologically protected localized 
modes~\cite{xiao2007valley}. 

\subsection{Dispersion Analysis of a Finite Strip and Transient Simulations}
\begin{figure}
	\centering
	\subfigure[]{
	\includegraphics[width=0.5\linewidth]{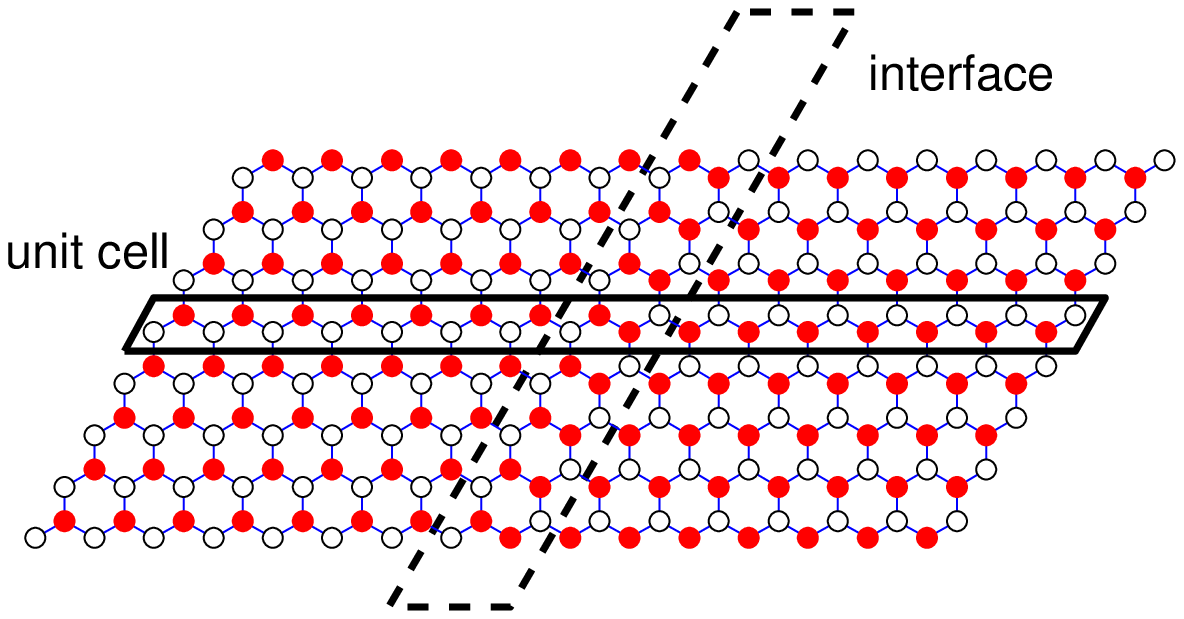}
	\label{fig:stripSchematic}
	}\\
	\subfigure[]{
	\includegraphics[width=0.45\linewidth]{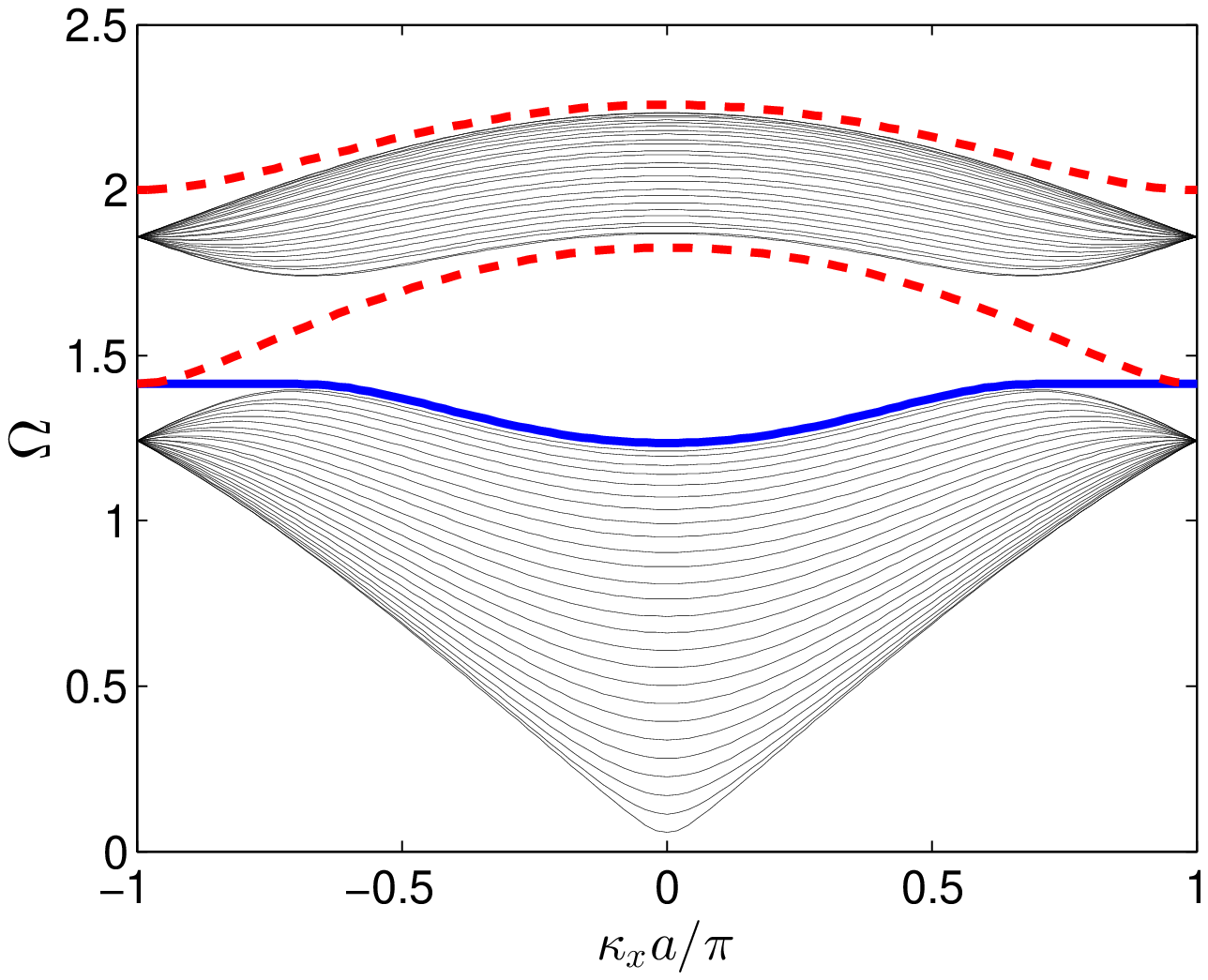}
	\label{fig:2D_type1Disp}
}
	\subfigure[]{
	\includegraphics[width=0.45	\linewidth]{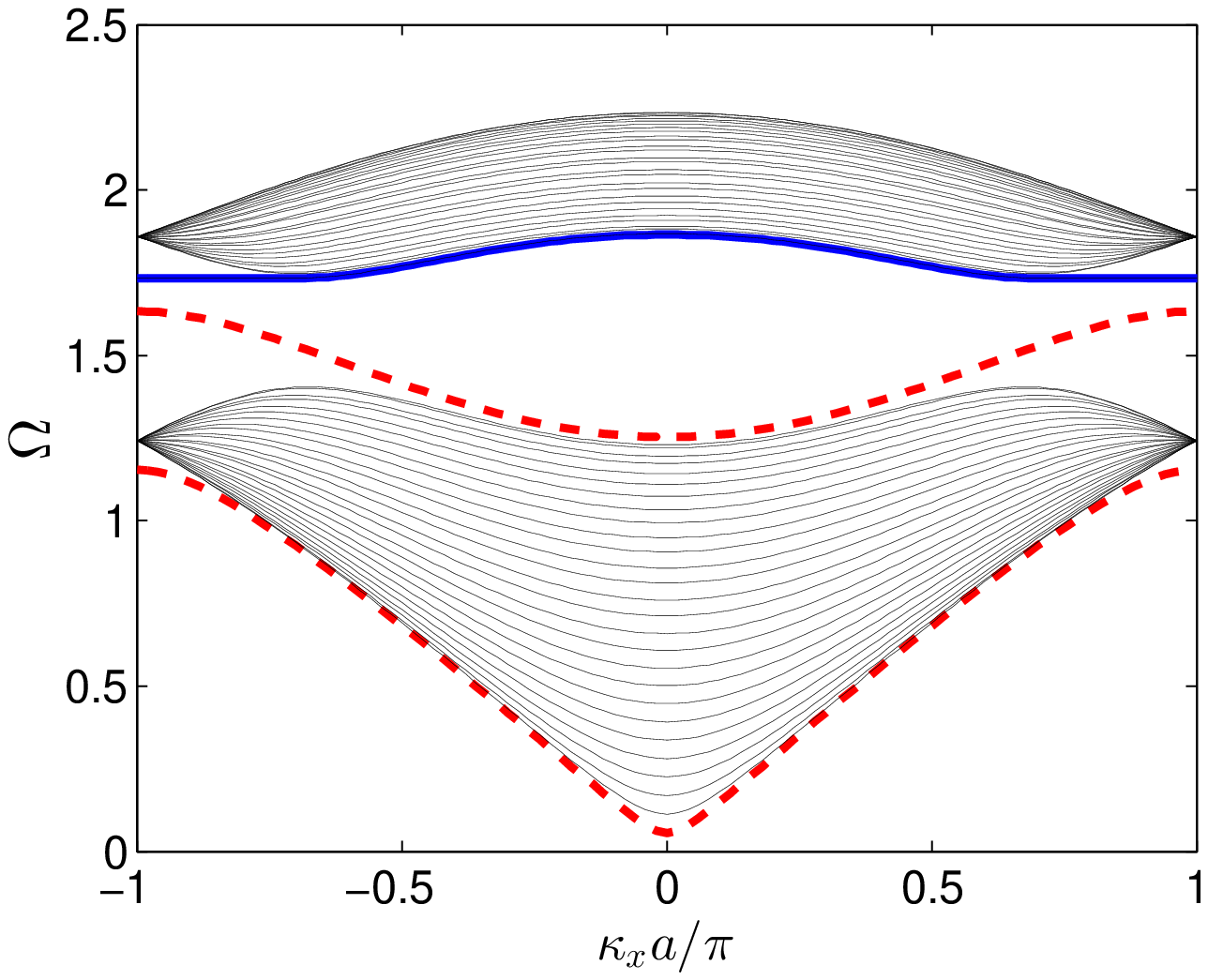}
	\label{fig:2D_type2Disp}
	}
\caption{(a) Schematic of unit cell considered for the dispersion analysis of a strip with an `L' interface (light masses are denoted as red, filled circles). (b) Dispersion diagram corresponding to an `L' interface and (c)  to an `H' interface (bulk modes - solid black lines, edge modes - dashed red lines). }
\end{figure}

Next, we analyze the dynamic behavior at an
interface between two hexagonal lattices with different unit cells. The unit cells  on one side of the interface have mass 
parameter $+\beta$ and the unit cells on the other side have mass parameter $-\beta$.
These two lattices thus have the same bulk band structure. Similar to the $1D$ case, we label a unit cell
of type $A$ or $B$ when $\beta>0$ or $\beta<0$, respectively. A strip, infinite along $\ba_2$ and finite along 
$\ba_1$ and consists of $16$ unit cells of each type is considered for our calculations. Figure~\ref{fig:stripSchematic} 
displays a schematic of part of the strip, along with the interface and a part of the unit cell.  
The interface is located along a line parallel to the $\ba_2$ direction, while the unit cell is a strip parallel to the $\ba_1$ direction as
sketched in the schematic. Note that the unit cell is periodic only along the $\ba_2$ direction. 
As in the 1D case, two types of interfaces can be constructed, with the interface having connecting two light or two heavy 
masses. These two interfaces are denoted as `L' and `H' respectively. An example of an `L' interface is illustrated in 
Fig.~\ref{fig:stripSchematic}, where the light masses are denoted as solid red circles, while the heavy masses are denoted as empty circles.

The band structure is evaluated by  fixing the masses at the left and right boundaries. 
Figure~\ref{fig:2D_type1Disp} displays the band diagrams for an `L' interface and for $m^a = 1.0$ and $m^b = 1.5$ (or equivalently, $m = 1.25$
and $\beta=0.25$). The diagram features two sets of bulk modes (black solid lines), along with two sets of modes within the bulk bandgap. 
These modes within the bandgap are localized modes either at the fixed ends or at the interface. The type of boundary where the mode 
localizes can be determined by examining the corresponding eigenvectors, which decay rapidly away from the boundary. 
The blue (thick solid) curve is associated with two overlapping frequencies corresponding to 
a localized mode at each end of the strip. Note that the two ends are locally identical, which results in this degeneracy of localized
modes.  
The branch denoted by the dashed red line is a single mode localized at the interface. In addition, a third branch is observed 
above the bulk optical band, 
illustrated in red dashed (light) color in the figure. The eigenvector associated with this mode is also localized 
at the interface. However, note that it is hard to excite
this mode in practice as the frequency it spans also has a wide spectrum of bulk bands.

We now consider a strip characterized by a `H' type interface, i.e. with two heavy masses adjacent to each other at the interface. 
The mass $m$ is kept the same, while $\beta$ is reversed in sign compared to the 
previous case. Figure~\ref{fig:2D_type2Disp} displays the dispersion diagram for this interface, having a different set of localized modes within the bulk bandgap. Here the mode localized at the interface, shown
again in dashed red (light) color starts from the bulk acoustics band, in contrast with the previous case. Furthermore, the two localized modes 
at the boundaries also have different frequencies, as there are now light masses at each boundary. Note also that there is an additional 
interface localized mode which has shifted downward below the acoustic band. 

\begin{figure}
	\centering
	\subfigure[]{
	\includegraphics[width=0.4\linewidth,clip=true]{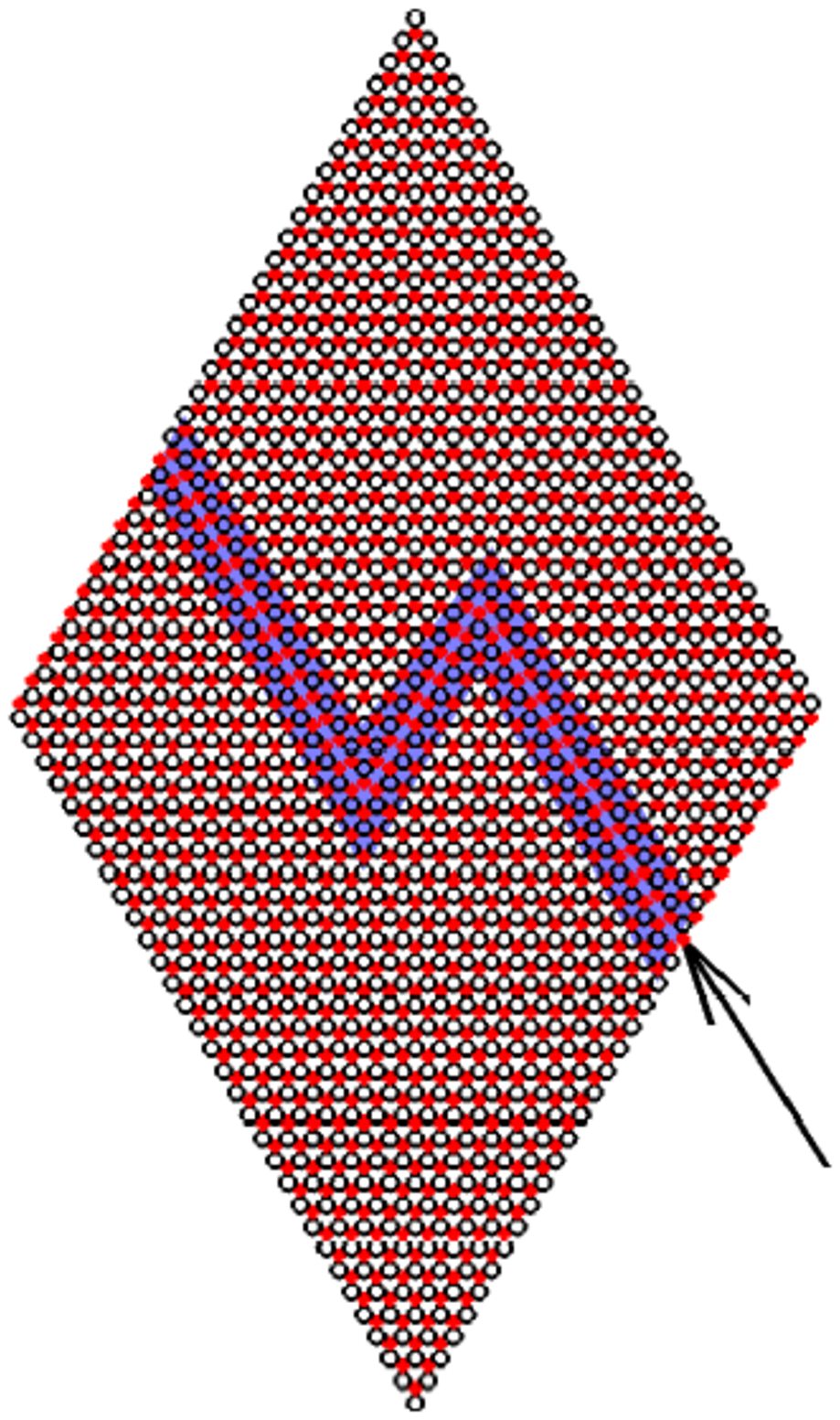}
	\label{fig:schemeFinite}
	}\\
	\subfigure[]{
	\includegraphics[width=0.3\linewidth,clip=true]{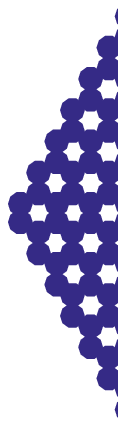}\label{fig:Ta}
	}
	\subfigure[]{
	\includegraphics[width=0.3\linewidth,clip=true]{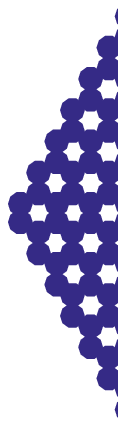}\label{fig:Tb}
	}
	\subfigure[]{
	\includegraphics[width=0.3\linewidth,clip=true]{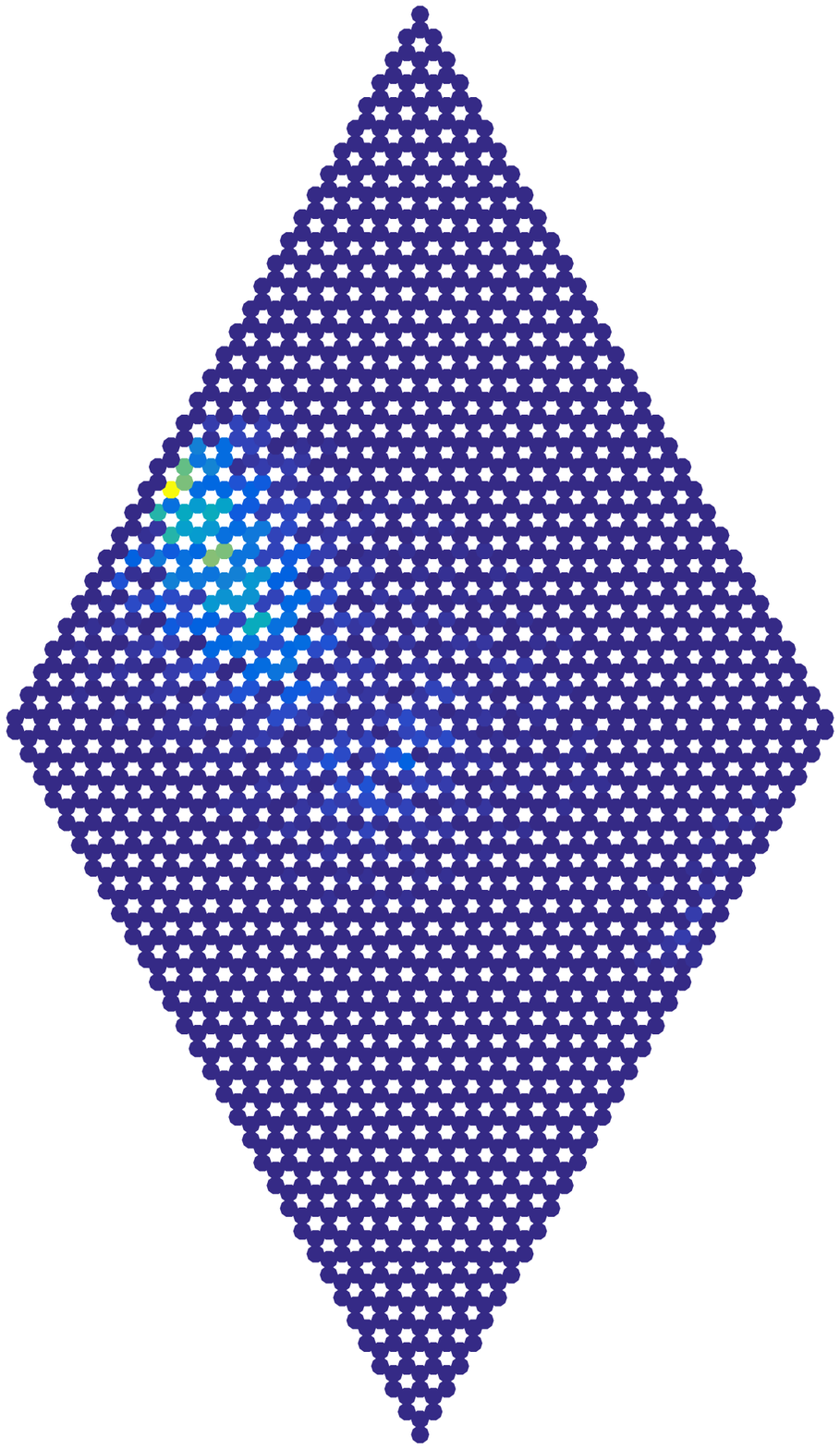}\label{fig:Td}
	}
	\caption{Transient simulation illustrating backscattering suppressed wave propagation along a zig-zag channel at $3$ 
	time instants: (a) Lattice schematic, (b) $t= 106$, (c) $t = 166$ and (d) $t = 240$.}
	\label{fig:2DdiscreteTransient}
\end{figure}

To verify the observations based on the above dispersion analysis, we conduct transient simulations on a finite lattice of $32\times 32$
unit cells. Both types $A$ and $B$ of unit cells are in the lattice, separated by a zig-zag interface of the `L' type (see Fig.~\ref{fig:schemeFinite}). The excitation is a $30$-cycles sinusoidal force of frequency $\Omega_e = 1.5$ modulated by a Hanning window and applied to one of the interface masses along the lower right boundary, shown by an arrow in the schematic. 
The response of the lattice is evaluated through numerical integration of the equation of motion for the finite system considered. Figure~\ref{fig:2DdiscreteTransient} displays the amplitude of the displacements in the lattice 
at $3$ distinct time instants. Initially, at $t = 106$ in Fig.~\ref{fig:Ta}, 
the wave travels along the straight portion of the interface and does not 
propagate into the interior or along the boundaries of the lattice. This solution is consistent with the dispersion analysis which predicts only a 
localized interface mode at frequency $\Omega_e$. As time progresses ($t = 166$ in Fig.~\ref{fig:Tb}), the wave bends around the zig-zag edges without any back-scattering. The wave is immune to localization and experiences negligible backscattering even as it navigates multiple bends as clearly shown from the displacement amplitude contours at 
$t = 240$ (Fig.~\ref{fig:Td}). After the wave hits the other boundary, it reflects and traverses in the opposite direction, as it is not immune 
to backscattering at the boundary, where mode hybridization occurs. We note that this response 
is consistent with the behavior observed for other classical analogues of the quantum Hall effect, which are also immune to backscattering only in the presence of a certain class of defects that do not cause the modes at the two valleys to 
hybridize. 

\section{Edge Waves in Continuous Elastic Plates}\label{sec:numerResult}
The concepts illustrated in the previous sections for discrete lattices are now extended to the case of an elastic plate carrying an array of resonators arranged in a hexagonal lattice topology. Thus, we seek to investigate the existence edge waves in a physical substrate, i.e. the elastic plates, that are described as a continuous elastic system. 

\begin{figure}
	\centering
	\subfigure[]{
	\includegraphics[width=0.65\linewidth]{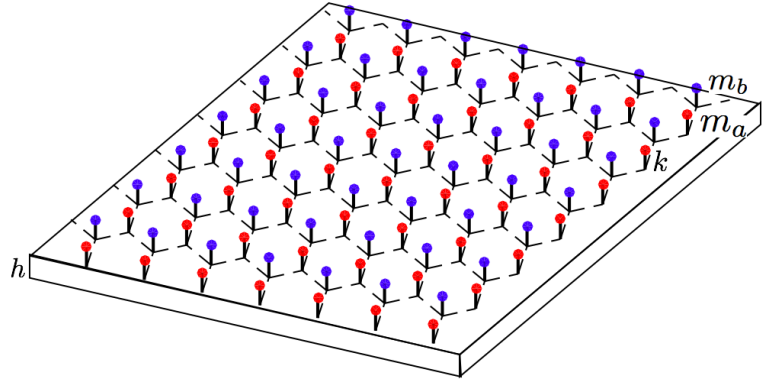}
	}\\
	\subfigure[]{
	\includegraphics[width=0.3\linewidth]{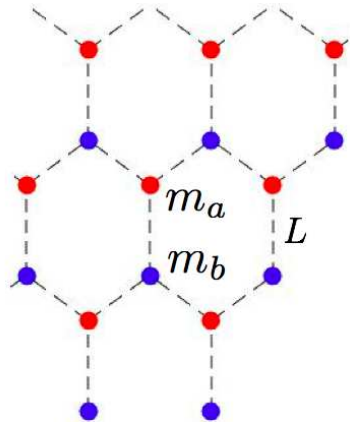}
	}
	\caption{(a) Schematic of a plate with resonators having two different masses and arranged in hexagonal lattice topology. (b) Detail of lattice configuration showing arrangement of different masses and nearest neighbor distance $L$.}
\label{fig:Plateschematic}
\end{figure}

The proposed configuration is based on the observation of Dirac cones in a plate with resonators placed in a hexagonal lattice arrangement~\cite{torrent2013elastic}. Breaking inversion symmetry within the unit cell can be simply achieved by considering
different masses for the two resonators. Figure~\ref{fig:Plateschematic} displays a schematic of the plate with two kinds of resonators, 
along with a zoomed-in view showing a unit cell with two resonators. The dashed lines are only to clearly depict the hexagonal geometry 
and are not associated with any physical features. 
Similar to the discrete case, breaking inversion symmetry while preserving $C_3$ symmetry leads to a bandgap in place of a Dirac point. 

\subsection{Plate Configuration and Governing Equations}
We consider a thin plate of Young's modulus $E$, Poisson's ratio $\nu$, thickness $h$ and density $\rho$. 
The resonators interact with the plate through a spring of stiffness $k$ and are located at positions $\bR_{\alpha}$. 
Also, let $a$ be the length of a unit cell of this hexagonal lattice, leading to a distance $L=a/\sqrt{3}$ between nearest neighbor resonators. 

Let $w(\bx)$ and $w_{\alpha}(R_\alpha)$ denote, respectively, the transverse displacement of the plate at point $\bx$ and the displacement 
of the resonator of type $\alpha$ located at $\bR_\alpha$.  The governing equations are then given by~\cite{xiao2012flexural} 
\begin{subequations}
\begin{gather}
D \nabla^4 w + \rho h \ddot{ w }  =  k \sum_{\alpha} \left( w - w_\alpha\right) \delta\left(\bx - \bR_\alpha\right),  \\ 
m^\alpha \ddot{w}_{\alpha}  + k\left( w_\alpha - w(\bR_{\alpha}) \right) = 0, 
\end{gather}\label{eqn:PlateRes} \end{subequations}
with $D = Eh^3/12(1-\nu^2)$ being the plate bending stiffness. Similar to the discrete case, the resonator masses at the two lattice
sites are expressed as $m^a = m (1 + \beta)$ and $m^b = m(1-\beta)$. 
For convenience, the following normalized frequency is introduced: 
\begin{equation*}
\Omega^2 = \omega^2 \dfrac{\rho a^2 h}{D}.
\end{equation*}
Also, the mass of the resonators is expressed in terms of the mass ratio $\gamma$, given by~\cite{torrent2013elastic}
\begin{equation*}
\gamma = \dfrac{m}{\rho A_c h},
\end{equation*}
where $A_c = \sqrt{3}a^2/2$ is the area of a unit cell. 
For all the calculations in this work, we use the resonator properties are obtained by considering $\gamma = 10$ and a baseline resonance frequency $\Omega_R a = a^2\sqrt{(k/m) \rho  h/D} = 4\pi$\cite{torrent2013elastic}. 

\subsection{Analysis of Plate Dispersion}

The dispersion properties of the plate are obtained through the application of the plane wave expansion method (PWEM)~\cite{xiao2012flexural}, which expresses the plate displacement $w(\bx,t)$ in terms of a finite number of orthogonal modes as
\begin{equation*}
w(\bx,t) = e^{ i(\omega t + \bkappa\cdot \bx )} \sum_{p,q = -M}^M e^{i (p\bg_1+ q\bg_2)\cdot\bx } \rm{w}_{p,q}(\bkappa), 
\end{equation*}
with $\rm{w}_{p,q}(\bkappa)$ being the coefficient associated with the mode $p,q$. 
Note that the basis vectors for the modes are reciprocal lattice vectors $(\bg_1,\bg_2)$ and are hereafter denoted by the set $G$, which contains  $N = (2M+1)^2$ terms of the form $\bm g=p\bg_1 + q\bg_2$. The displacement field may then be written as 
$w(\bx,t) = e^{i\omega t + i\kappa\cdot \bx}\sum_{G} e^{i\bg \cdot \bx}\rm{w}_G$.
Similarly, using Floquet Bloch theory, the displacement of a resonator located at $\bR_{\alpha}$ can be expressed
as $w_{\alpha}(\bR_\alpha) = \sum_{\bkappa} e^{i\bkappa \cdot \bR_\alpha} \rm{w}_{\alpha}(\bkappa)$. 
Substituting these expressions into the plate governing equation and using the normalizing variables defined above 
leads to the following equation
\begin{gather}
\sum_{G'} \left( |\bkappa + \bg'|^4 - \Omega^2 /a^2 \right) e^{i(\bkappa + \bg') \cdot \bx} \rm{w}_{G'} = 
				e^{i \bkappa \cdot \bx}  \sum_{\alpha} \dfrac{k}{D}\left( w_{\alpha} - \sum_{G'}e^{i\bm {g'} \cdot \bx}\rm{w}_{G'} \right)\delta(\bx-\bR_{\alpha})  
\end{gather}
Multiplying by $e^{-i\bg \cdot \bx -i \bkappa \cdot \bx}$, integrating over one unit cell and 
applying orthogonality gives
\begin{equation}
\left( |\bkappa + \bg|^4 - \Omega^2/ a^2 \right) \rm{w}_G = 
				\dfrac{\gamma \Omega_R^2}{a^2}\sum_{\alpha}   e^{-i\bg \cdot\bR_{\alpha}} \left(\rm{w}_\alpha - \sum_{G^\prime} e^{i\bm{g'} \cdot \bR_{\alpha}}\rm{w}_{G'} \right). \label{plateEig} 
\end{equation}
Note that the index $\alpha$ takes values
$1$ and $2$ corresponding to the two resonators within a unit cell.
Similarly, substituting the displacement fields in Eqn.~\eqref{eqn:PlateRes} for the resonator gives:
\begin{equation}
-\Omega^2 (1 - (-1)^\alpha \beta) \rm{w}_\alpha + \Omega_R^2 \rm{w}_{\alpha} - \Omega_R^2 \sum_G e^{i\bg \cdot \bR_{\alpha}} \rm{w}_G  = 0,
\label{plateMassEig}
\end{equation}
Equations~\eqref{plateEig} and~\eqref{plateMassEig} define an eigenvalue problem for every  
wave vector $\bkappa$. The resulting eigenvalues and eigenvectors yield the frequency $\Omega$ and the associated displacement field.

\begin{figure}
\centering
\includegraphics[width=0.6\linewidth]{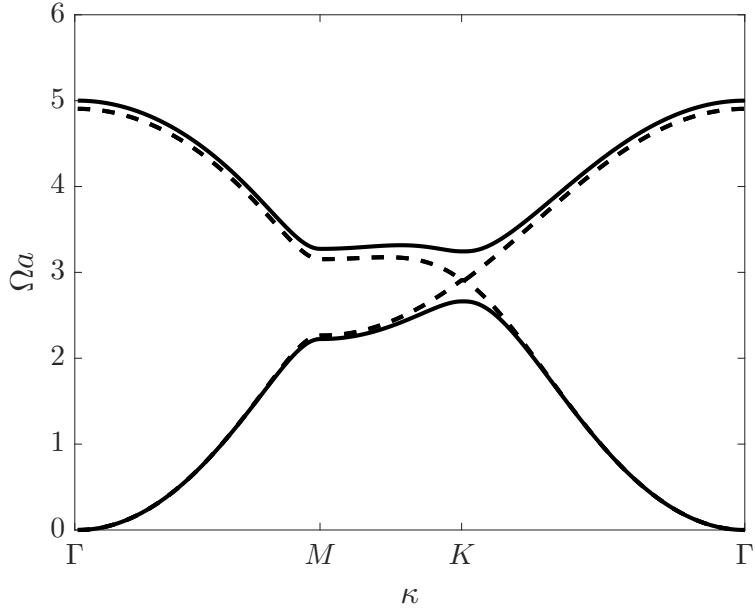}
\caption{Dispersion curves over the IBZ of a unit cell with equal (dashed curves) and unequal (solid curves) masses, showing the Dirac point and the formation of a bandgap with unequal masses.}
\label{fig:IBZ_plateRes}
\end{figure}

Figure~\ref{fig:IBZ_plateRes} displays the plate dispersion diagram for two kinds of resonator unit cells. The dashed curves show the frequency for a unit cell having both masses equal $\beta=0$, while the solid curves are the dispersion diagrams for a unit cell having dissimilar masses in the resonators ($\beta=-0.4$). As shown by Torrent and coworkers~\cite{torrent2013elastic}, when the resonator masses are identical, there is a Dirac cone at the $K$ point. In contrast, when the two masses are different, a bandgap forms, which essentially replicates the behavior of the discrete 2D lattice previously examined.

The nontrivial  nature of the bandgaps is characterized by the valley Chern number, which is calculated by a similar approach 
used for the discrete lattice. 
To this end, the eigenvalue problem defined by Eqns.~\eqref{plateEig} and~\eqref{plateMassEig} is expressed as 
$\Omega^2 \bM \bv =\bK \bv$, where $\bv = [\rm{w}_G;\rm{w}_\alpha]$ is a generalized displacement eigenvector 
and where $\bM$ and $\bK$ are the generalized mass and stiffness matrix operators, given by
the coefficients of $\Omega^2$ and $\Omega^0$, respectively, in these equations.
Although the number of bands obtained in the solution depend on the number of terms used in the 
plane wave expansion, the frequencies of the first two bands are well separated from the remaining bands and 
their contribution to the Berry curvature 
is negligible as evident from the denominator term of Eqn.~\eqref{eqn:BerryCurv}. Thus, only the two lowest bands having
eigenvalues $(\Omega_m,\Omega_n)$ and associated eigenvectors $(\bmm,\bn)$ are used for 
calculating the Berry curvature and Eqn.~\eqref{eqn:BerryCurv} for the first band having eigenvector $\bmm$ reduces to 
\begin{equation}
B(\bkappa) = \dfrac{ \Braket{\bmm | \dfrac{\partial \bK}{\partial \kappa_x} |  \bn}
\Braket{\bn | \dfrac{\partial \bK}{\partial \kappa_y} | \bmm}  - c.c.}{\left( \Omega^2_m - \Omega^2_n \right)^2}. 
\end{equation}

To evaluate the derivatives of the generalized stiffness $\bK$ 
with respect to the wave vector components, observe that only the first term in Eqn.~\eqref{plateEig} depends on $\kappa$ and all other
terms in the eigenvalue problem (Eqns.~\eqref{plateEig} and~\eqref{plateMassEig}) are independent of the wave-vector.
Thus the derivative of the stiffness matrix $\bT = {\partial \bK/\partial k_x}$ has nonzero components only due to the $N\times N$ diagonal 
terms $|\bkappa + \bg|^4$, arising from the $N\times N$ reciprocal lattice vectors $\bg$. In particular, the nonzero component of $\bT$ 
associated with the $p$-th component of $\rm{w}_G$ is 
$T(p,p) = 4(k_x + g_x(p))( (k_x + g_x(p))^2 + (k_y + g_y(p))^2)$. 
Using this expression and the solution of the eigenvalue problem
which yields the eigenvalues of the first two bands $(\Omega_m,\Omega_n)$ and their associated eigenvectors, the Berry curvature associated
with each band over the entire Brillouin zone can be evaluated. 

\begin{figure}
	\centering
	\subfigure[]{
	\includegraphics[width=0.45\linewidth]{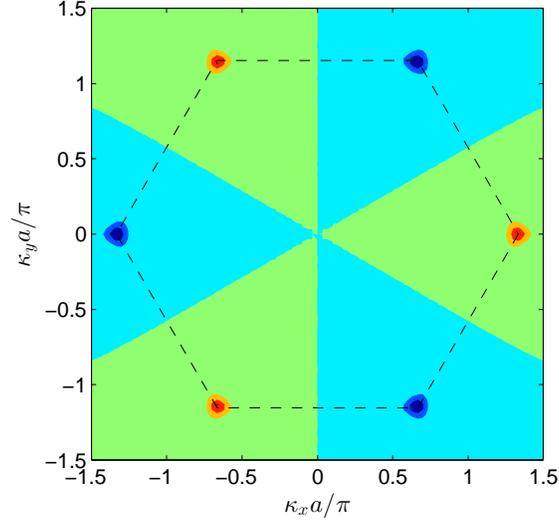}
	\label{fig:BerryCurv}
	}\\
	\subfigure[]{
	\includegraphics[width=0.6\linewidth]{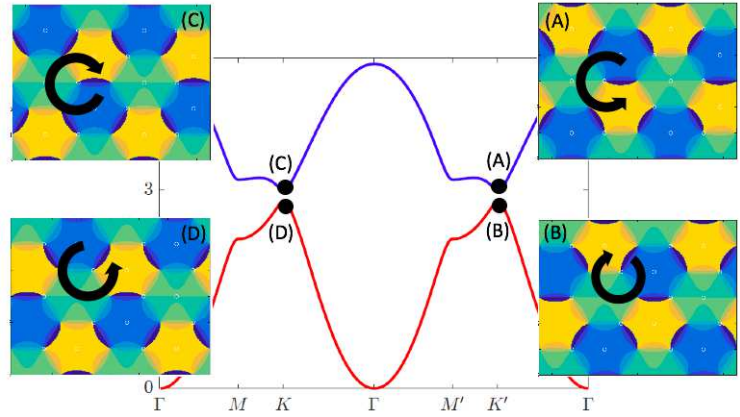}
	\label{fig:ValleyModes}
	}
	\caption{(a) Berry curvature over the reciprocal lattice space for the acoustic branch. 
It is localized at the $K$ and $K'$ valley points and has opposite signs at these points. (b) Phase of the eigenmodes at the 
$K$ and $K'$ points have opposite polarization.}
\end{figure}

Figure~\ref{fig:BerryCurv} displays the Berry curvature associated with the first band for a plate with $\beta=-0.1$ and it 
is localized near the $K$ and $K'$ points. Similar to the $2D$ discrete case, where it also has opposite signs at the $K$ and $K'$ points. The valley Chern number associated with these two valleys are 
$+1/2$ and $-1/2$, respectively. Similarly, for the second band, the valley Chern numbers at the $K$ and $K'$ points are flipped in sign as
the sum of Chern numbers over the two bands is zero. The nontrivial value of Chern numbers thus predicts the presence of topologically 
protected valley edge modes at the interface between two lattices whose corresponding valleys have opposite Chern numbers.  
Two such lattices are constructed by simply flipping the values of the two masses $m^a$ and $m^b$ leading to a different kinds of unit 
cell. Thus at the interface between two lattices, one having $\beta>0$ (or $m^a > m^b$) and the other having $\beta<0$ (or $m^a < m^b$)
unit cells, topologically protected valley
edge modes exist due to the nontrivial topology associated with the bands. For additional insight into the nature of the eigenmodes at the valley points, we examine their phases. Figure~\ref{fig:ValleyModes} displays the dispersion diagram over
the $\Gamma-M-K-\Gamma$ and the $\Gamma-M'-K'-\Gamma$ boundaries of Brillouin zone. 
The insets show the phase of the displacement field for each of the eigenmodes at the $K$ and $K'$ points. The polarization plots show that although the eigenvalues are identical,  the eigenmodes have opposite polarizations at the $K$ and $K'$ points for each band. 
Similar polarization reversal is observed in photonic crystals~\cite{chen2016valley}. 



\subsection{Finite Strip Dispersion Analysis and Finite Lattice Simulations}
The predictions from the dispersion analysis above are verified through the dispersion analysis extended to a finite strip with an interface, along with simulation results of the response of finite plates. The dispersion diagrams for a strip are presented, similar to the previous discrete case, followed by 
multiple scattering simulations of a finite lattice which illustrate wave transmission along an interface even  in the presence of defects. 

For convenience in dispersion calculations on a  strip, 
the lattice vectors are chosen to be orthogonal, given by $\ba_1 = a[1,0]$ and $\ba_2 =\sqrt{3}a[0,1]$. Each lattice
unit cell has $4$ resonators for this choice of lattice vectors. A strip of $N=8$ unit cells along the $\ba_2$ direction
is considered for the study. The length of the strip is chosen to be
$H = (N + 2)\sqrt{3}a $ in the vertical direction. Note that we have chosen the plate to be one unit cell longer than the zone of
resonators on either side. 

Consider the set of  resonator displacements in the lattice described by $w_{\alpha}(\bx,t)$, 
where the index $\alpha$ runs from $1$ to $2M$ for all the resonators in the strip $(x,y) \in [0, a]\times[0 ,H]$. 
A plane wave solution in the $\bm a_1$-direction is imposed of the form
\begin{equation}
w(\bx,t) = e^{i (\kappa_x x + \omega t) } \sum_{p} e^{i  g_p x}\rm{w}_m(y) 
\end{equation}
for the plate, where $g_m =  2\pi m/a, \;m\in \{-M,...,M\}$. Note that in contrast with the 
infinite lattice, only the $x$-direction is periodic, and the goal is to obtain the corresponding set of functions $\rm{w}_{m}(y)$ from an eigenvalue problem for assigned $\kappa$. 
Substituting it into the governing equation, multiplying both sides by $e^{-i(g_m+\kappa)\cdot x}$ and integrating over
the strip $[0,a]\times[0,H]$ leads to the following equation for each 
reciprocal lattice vector $g_m$ and wavenumber $\kappa_x$
\begin{equation}\label{eqn:FEA1}
\left( \dfrac{d^4}{dy^4} + 2(\kappa_x + g_p )^2 \dfrac{d^2 }{dy^2} +  (k_x + g_p )^4 - \omega^2 \dfrac{\rho h}{ D}\right)\rm{w}_m = 
			\dfrac{\sqrt{3}\gamma \Omega_R^2}{2 aH}\sum_{\alpha} e^{-i g_p x_\alpha}\left( \rm{w}_{\alpha} - \sum_{p} e^{i g_p x_\alpha}	\rm{w}_{n}(y_{\alpha}) \right). 
\end{equation}
Similarly, the resonator equation becomes 
\begin{equation}\label{eqn:FEA2}
-\Omega^2  (1\pm \beta) \rm{w}_{\alpha} = \Omega_R^2 \left( \sum_{n}\rm{w}_n (y_{\alpha})e^{i g_p x_{\alpha} } - \rm{w}_{\alpha} \right),
\end{equation}
with the first term within brackets taking values $1+\beta$ or $1-\beta$ depending on the resonator type. 
The above system of equations define an eigenvalue problem with eigenvalue $\omega$ and 
eigenvector having $2M+1$ functions $\rm{w}_{m}(y)$. It is solved numerically using 
beam finite elements using the procedure described in 
the appendix~\ref{sec:app1}.

\begin{figure}
	\centering
	\subfigure[]{
	\includegraphics[width=0.45\linewidth]{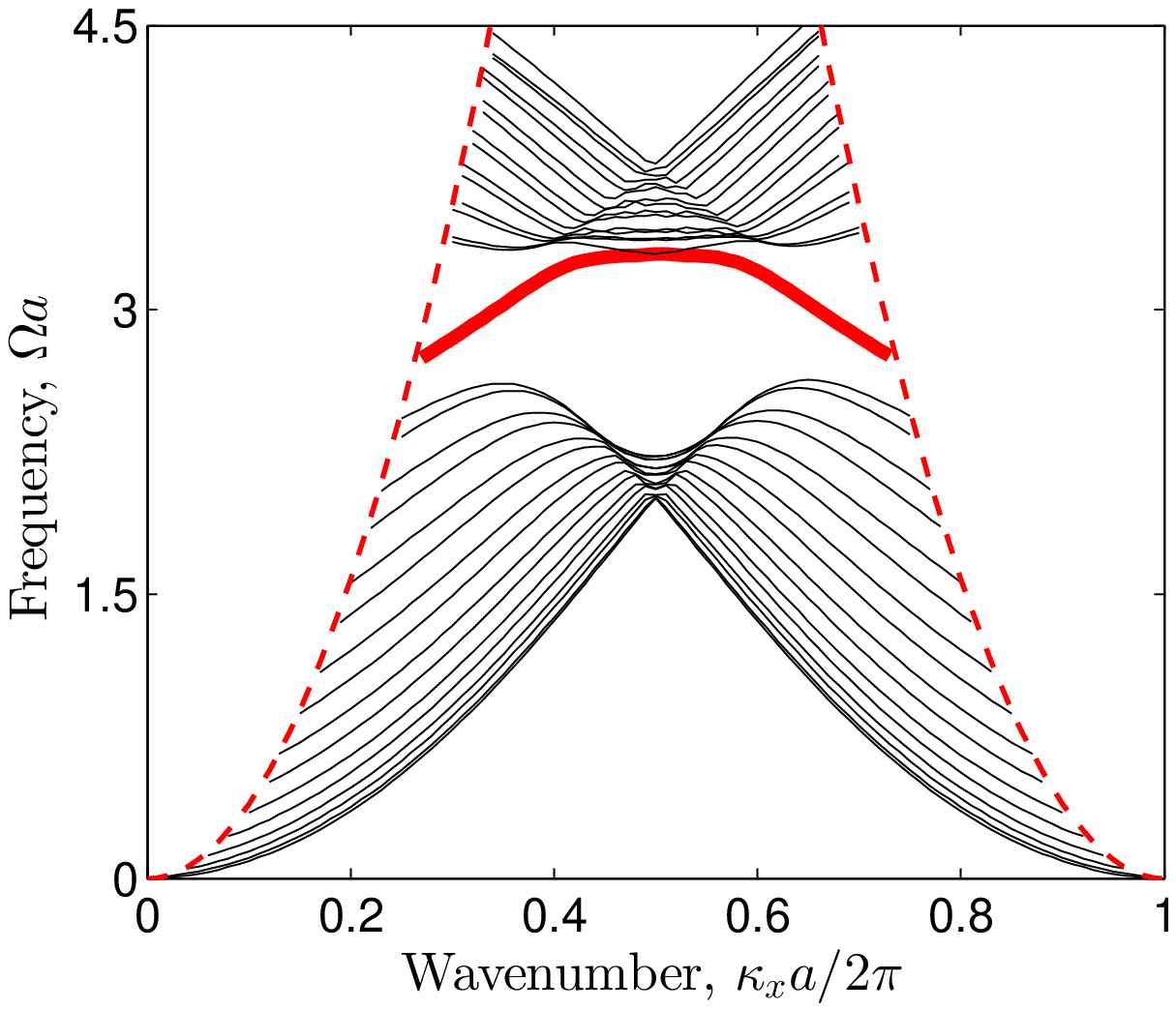}
	\label{fig:plateStripDisp_1}
	}
	\subfigure[]{
	\includegraphics[width=0.45\linewidth]{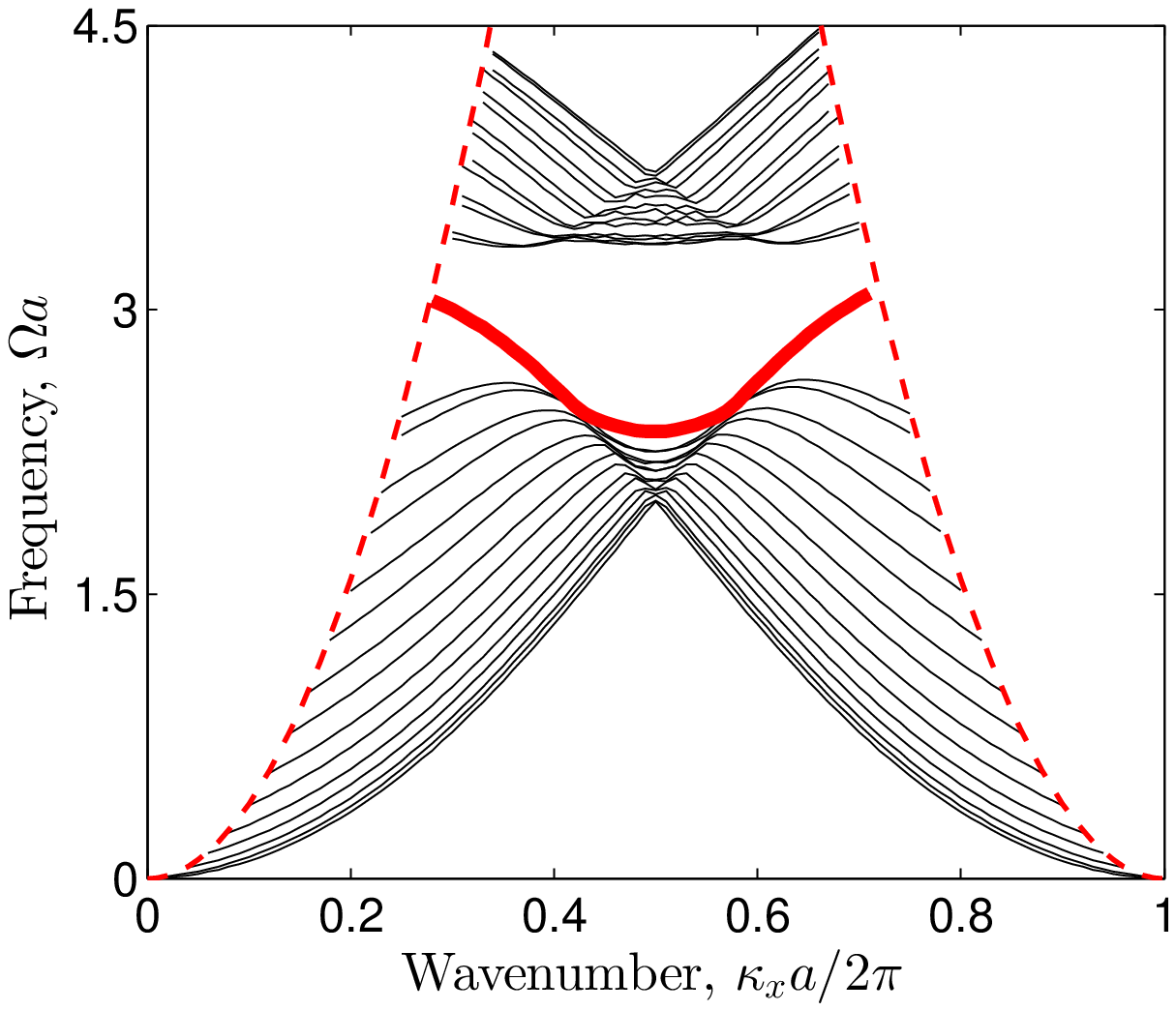}
	\label{fig:plateStripDisp_2}
	}
	\subfigure[]{
	\includegraphics[width=0.45\linewidth]{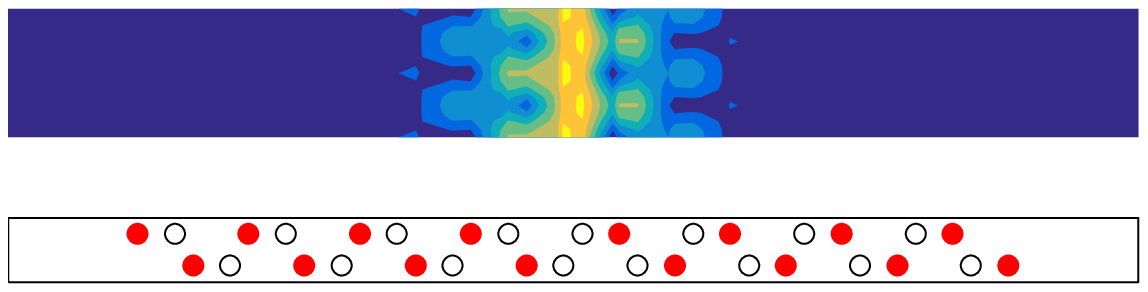}
	\label{fig:eigVec_Plate_Config1}
	}
	\subfigure[]{
	\includegraphics[width=0.45\linewidth]{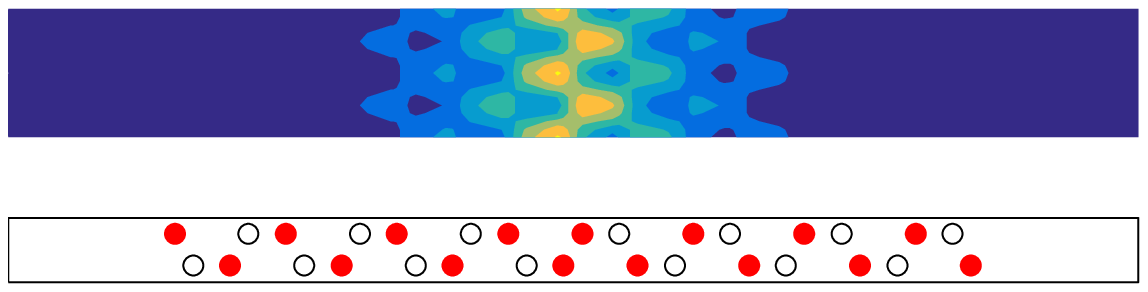}
	\label{fig:eigVec_Plate_Config2}
	}
	\caption{Dispersion diagram  and localized modes 
	for a plate having a finite width strip of resonators for two types of interface. Lattice configuration with adjacent
  (a) light and	(b) heavy masses at the interface. (c) and (d): corresponding mode shapes along with unit cell schematic. 
	}
	\label{fig:plateStripDisp}
\end{figure}

We now consider a strip of $N=8$ unit cells with $\beta=0.2$ 
and two types of lattice configurations, which 
result in distinct kinds of lattice interface, similar to the $2D$ discrete lattice case. 
Again, by virtue of bulk-boundary correspondence principle, topologically protected edge modes are guaranteed to exist at the interface between the two lattice types which have distinct valley 
Chern numbers. 
Note that the dispersion 
relation of a plate is $\omega^2 = \kappa_x^2 + \kappa_y^2$. For a waveguide along the $x$-direction, the allowable frequencies are 
$\omega \leq \kappa_x^2$ since $\kappa_y=0$. In the presence of 
resonators, this relation becomes periodic with a period $\pi/a$ in $\kappa_x$, which is indicated by the dashed red curves in the figure. 
Thus only the range of frequencies enclosed between these two dashed lines are admissible for waveguiding along $x$-direction in a plate. 

We first consider the case where the lattice is in a similar configuration to an interface denoted as `L' in the discrete case, 
with $4$ unit cells of 
each lattice type in a strip. Figure~\ref{fig:eigVec_Plate_Config1} displays a schematic of the unit cell. It has 
two adjacent resonators with light masses at the interface, separated by a distance $a/\sqrt{3}$. 
Figure~\ref{fig:plateStripDisp_1} displays the corresponding dispersion diagram and it has two sets of bulk bands 
with an edge mode between them. 
The edge mode present in the bandgap frequencies is associated with an eigenvector localized at the interface and is illustrated by 
its displacement amplitude contours in Fig.~\ref{fig:eigVec_Plate_Config1}, here calculated for $\kappa = 0.4 \pi/L$.

Figure~\ref{fig:plateStripDisp_2} displays the dispersion diagram for the strip having the an interface `H', i.e. with two adjacent 
heavier masses separated by a distance $a/\sqrt{3}$. 
Similar to the previous case, we observe a mode in the bulk bandgap. 
The frequency of this mode has a local minimum at $\kappa_x=\pi/a$. Note that this mode is different from the localized mode at the interface between adjacent light masses. Indeed, comparing the frequencies associated with these two interface localized modes, we see that the `L' interface is characterized by a localized mode spanning between the $K$ 
and $K'$ valleys in the optical band and has a peak frequency at $\kappa_x = \pi/a$, while in the `H' interface, the band
associated with the localized mode spans the two valleys in the acoustic band. This feature is similar to the valley edge modes observed by
Ma and Shvets~\cite{ma2016all} in photonic crystals. In contrast with the discrete case, only a light mass at the end (last resonator) 
results in a localized
defect mode at the boundary. The boundary has been modified here to have only heavy masses at both ends and so that no modes are localized at the domain outer boundary. 
The eigenmodes corresponding to the edge modes in the two lattice configurations are displayed in Figs.~\ref{fig:eigVec_Plate_Config1} 
and~\ref{fig:eigVec_Plate_Config2} at wavenumber $\kappa = 0.4 \pi/L$. 
The displacement amplitude in the plate strip are localized at the interface and verify our assertion of an interface mode.  
Note that the two modes are different and the amplitude is localized at the resonator having lower mass. 

Since these modes at the interface arise due to bands on either side of the interface having
distinct topological indices (valley Chern number), they are immune to backscattering and localization in the presence of a class
of defects, which do not cause hybridization of the two distinct valley modes. In contrast, the localized modes present at the boundary of the 
strip are defect modes which lack topological protection and are susceptible to localization at corners and defects. 
We support these assertions through numerical simulations on an infinite plate featuring a finite $20\times 20$ resonator array. Multiple scattering simulations are conducted following the procedure described in~\cite{torrent2013elastic} to predict displacement field in the plate resulting from a  point source excitation. Note that the method yields the steady state solution after all the transients have died out. 

\begin{figure}
	\centering
	\includegraphics[width=0.55\linewidth]{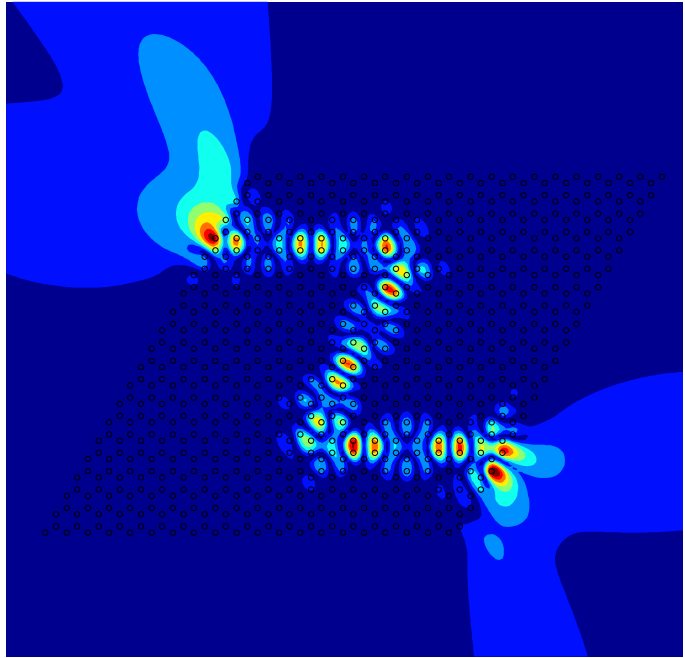}
	\caption{Steady state displacement field in a plate having Z-shaped interface between the two unit cell types  
	The plate is excited with a point source located on the left edge . The wave propagation is 
	confined to the interface and does not localize at the corners. 
	}
	\label{fig:plateZchannel}
\end{figure}

Figure~\ref{fig:schematicPlate} displays a schematic of the problem setup with two types of unit cells in the lattice 
separated by a $Z$-shaped interface. The unit cells below the interface are of type $A$, with $\beta> 0$, while those above the interface
have $\beta < 0$ and there are two adjacent light masses at the interface. The results are presented here for a lattice with mass 
parameter $\beta=0.4$. 
Figure~\ref{fig:plateZchannel} illustrates the contours of displacement amplitude 
in the plate due to a point source placed on the left edge at the boundary 
between the two unit cell types and oscillating at a normalized frequency $\Omega a = 3.0$. The results clearly illustrate how the displacement is localized along the $Z$-shaped interface. The surrounding plate without resonators does not have a bandgap, and allows the energy to leak from the two ends of the channel into its unbounded domain. Note that there is no localization of energy even as the wave bends around corners. Furthermore, the wave amplitude is almost identical at the left and right boundary, which shows that there is no significant backscattering. 

\begin{figure}
	\centering
	\subfigure[]{
	\includegraphics[width=0.46\linewidth]{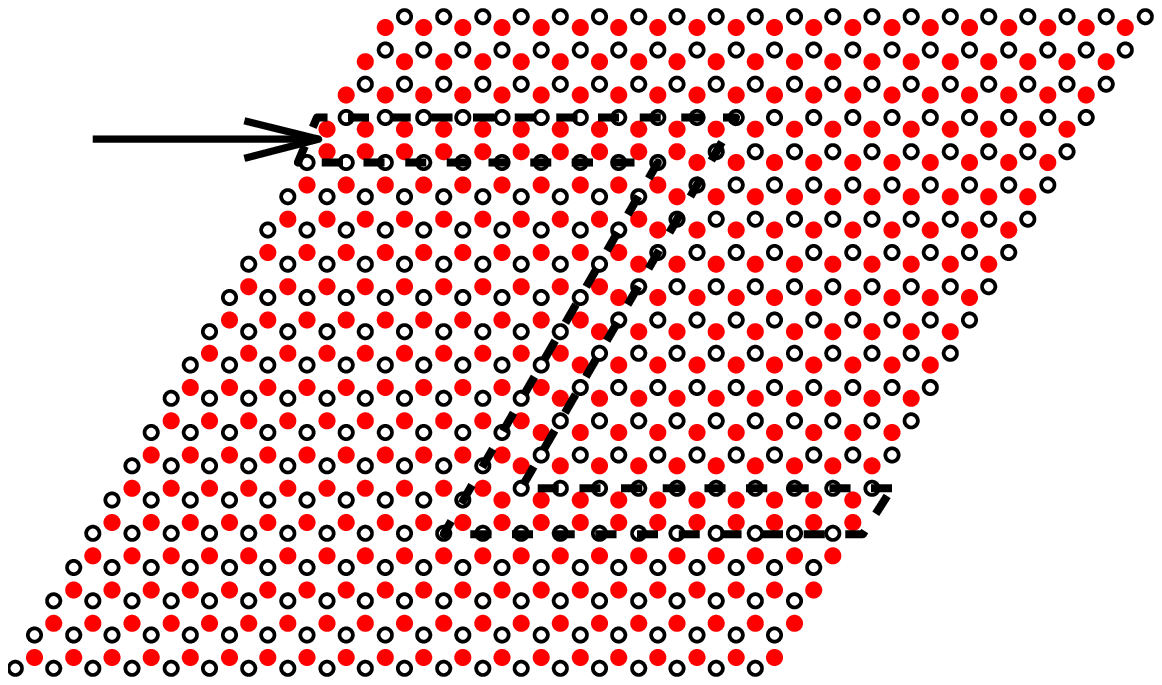}
	\label{fig:schematicPlate}
	}
	\subfigure[]{
	\includegraphics[width=0.46\linewidth]{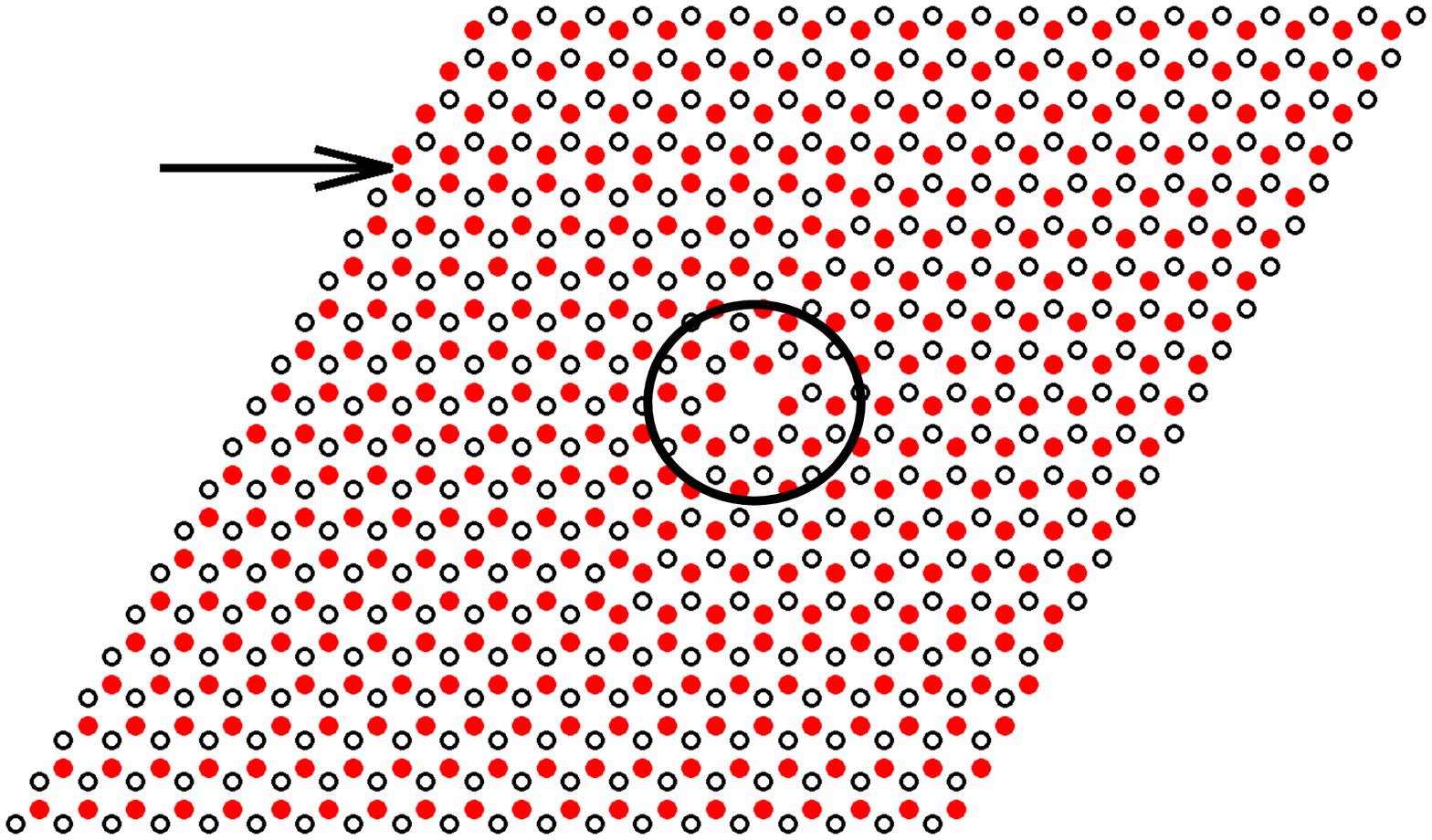}
	\label{fig:schematicPlate_Defect}
	}
	\caption{Schematic of a finite lattice having plate with resonators. The two types of unit cells are separated by a Z-shaped interface and a
	point excitation, shown by the arrow is applied on the left edge. (a) Plate with only corner defects and (b) plate with both corner 
	and vacancy defect, with a pair of resonators along the interface removed.  
	}
	\label{fig:schematicPlate}
\end{figure}

\begin{figure}
	\centering
	\subfigure[]{
	\includegraphics[width=0.45\linewidth]{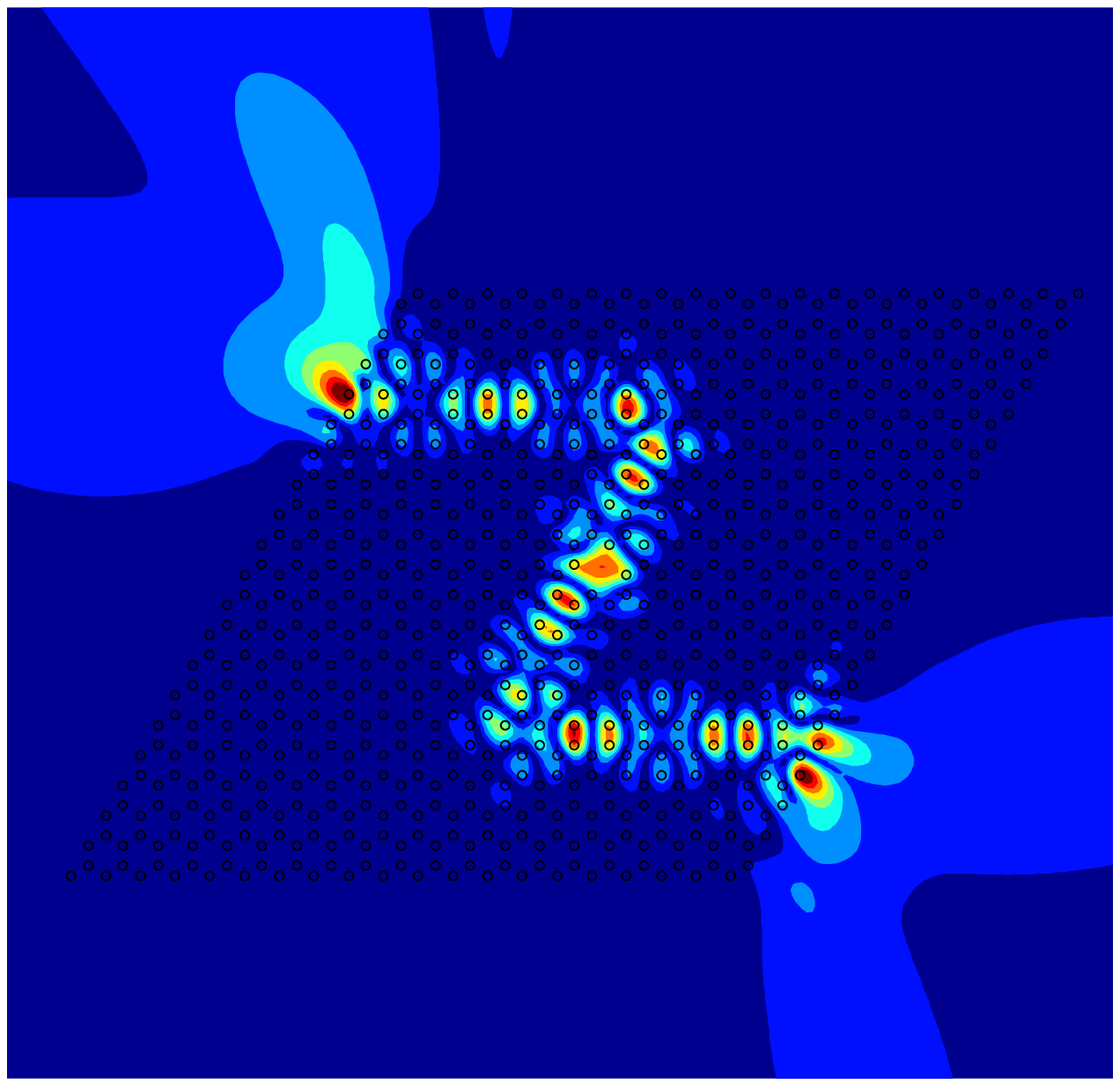}
	\label{fig:scatterGap}
	}
	\subfigure[]{
	\includegraphics[width=0.45\linewidth]{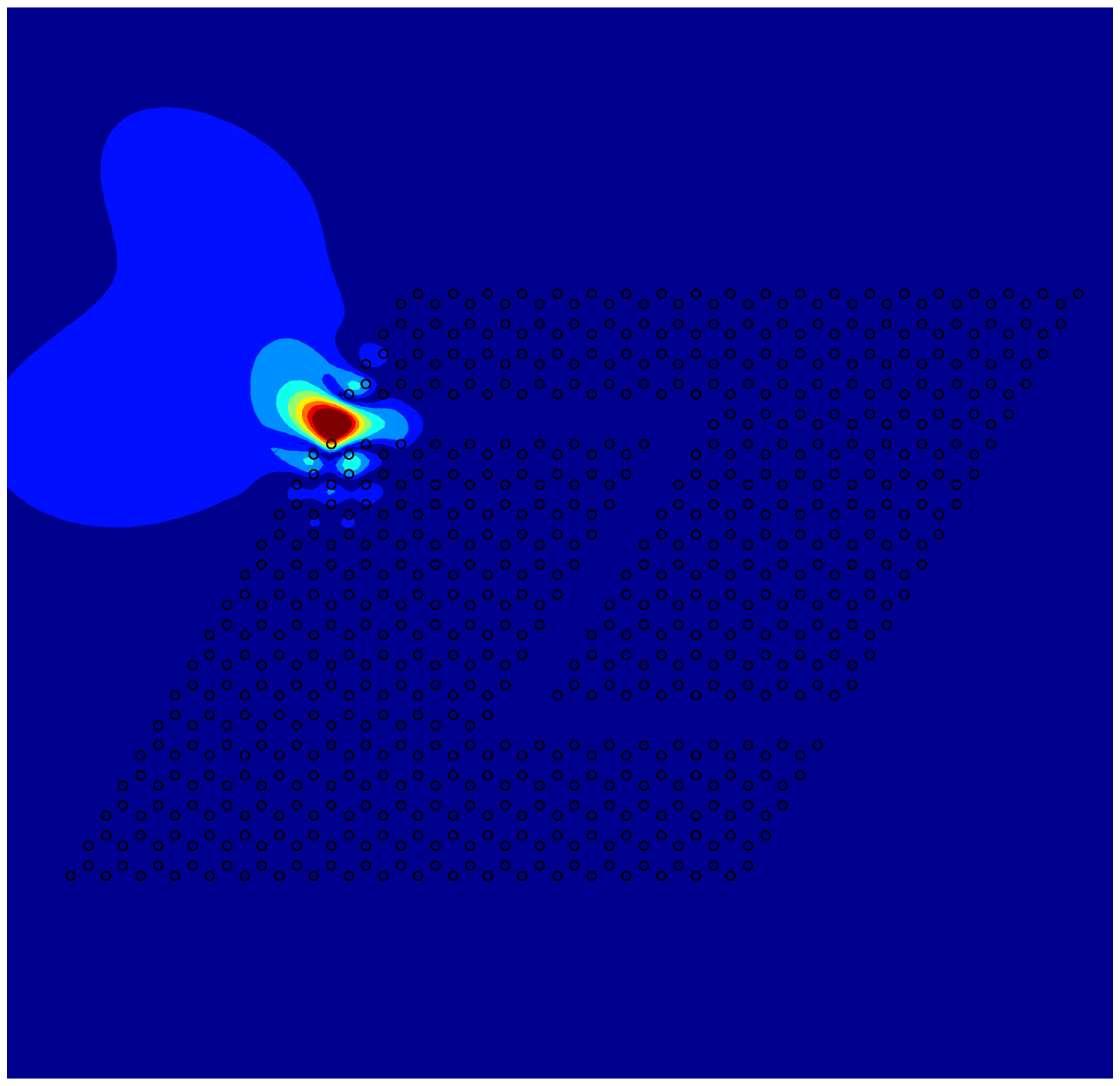}
	\label{fig:scatter2}
	}
	\caption{Effect of defects on wave propagation. (a) A point defect is introduced along the interface between two distinct 
	unit cell types by removing $2$ resonators
	from the center of the plate. (b) A line defect in a lattice with all identical unit cells. Localization occurs only in the second case.}
	\label{fig:plateDefects}
\end{figure}

We now show the results for two further kinds of lattice imperfections. The first lattice imperfection involves removing both the 
resonators from a single unit cell. Figure~\ref{fig:schematicPlate_Defect} displays a schematic of this plate with defects. It is identical to the previous case, except the removal of two resonators lying in the middle section of the Z-shaped interface. 
Figure~\ref{fig:scatterGap} illustrates the amplitude of the displacement field 
under a similar external excitation condition, with a point source placed at the interface on the left edge. We observe a robust propagation 
of waves even in the presence of this defect, thereby demonstrating immunity of the interface mode. Note that in contrast, in an 
ordinary bandgap in a lattice, removal of a unit cell may result in localized modes or it may induce strong backscattering effects.  
An example of such localization is shown by results obtained for a second lattice imperfection, which considers a lattice made of 
the same kind of unit cells ($\beta<0$) 
with a layer of unit cells removed along the same Z-shaped strip. Figure~\ref{fig:scatter2} illustrates the resulting displacement field which is strongly localized near the source and does not propagate into the channel. This localization happens because the defect is not wide enough compared to the wavelength associated with this frequency, causing the mode to hybridize with the evanescent mode in the direction
normal to the channel. 

\section{Conclusions}\label{sec:conc}

This study illustrates how breaking parity or  inversion symmetry within the 
unit cell can lead to topologically non-trivial bandgaps in both $1D$ and $2D$ mechanical 
lattices. These lattices are solely made of passive components and are characterized by a behavior which is analogous to the quantum valley Hall effect, whereby
difference in valley Chern numbers at various points in the Brillouin zone are pursued and exploited. The discrete lattices are extended to an elastic plate having resonators arranged in a hexagonal array. Similar ideas of breaking inversion
symmetry while preserving $C_3$ symmetry within the unit cell through proper choice of the resonators leads to non-trivial bandgaps in the plate. 

Theoretical predictions are verified through both dispersion analyses on extended unit cells and numerical simulations on finite lattices. 
The dispersion studies on finite strips show the presence of edge modes localized at the interface between two distinct lattice 
unit cell types. 
Transient simulations on the discrete hexagonal lattice show wave propagation along predefined interfaces even in the presence of multiple bends. 
Finally, multiple scattering simulations illustrate robust edge wave propagation in plates with resonators at a Z-shaped interface even in 
the presence of defects.

\section*{Acknowledgments}
This work is supported by a from the Air Force Office of Scientific Research (Grant Number: FA9550-13-1-0122).

\appendix
\section{Finite element formulation for plate strip dispersion analysis}\label{sec:app1}

The system of equations given by Eqns.~\eqref{eqn:FEA1} and~\eqref{eqn:FEA2} define an eigenvalue problem with eigenvalue $\omega$ and 
eigenvector having $2M+1$ functions $\rm{w}_{m}(y)$. The solution is conducted by discretizing the above equations through two-node Hermitian finite elements~\cite{cook2007concepts}. The shape functions $N(y)$ are chosen to be 
localized about the resonators  to evaluate the force on resonators accurately. Note that they satisfy a
partition of unity rule, i.e., $N_m(y_n) = \delta_{mn}$, where $\delta_{mn}$ is the Kroneker delta product. 
The solution field is expressed in terms of the displacement values of the $M$ degrees of freedom as 
\begin{equation*}
\rm{w}_n(y) = \sum_{r } \rm{w}_n(y_r) N_r (y).  
\end{equation*}
Substituting the above equation into the variational form of the plate equations leads to an eigenvalue problem in algebraic form, which is here solved for $M=1$, with $n g$ thus taking three values. The matrix form of this eigenvalue problem may then be expressed as 
\begin{equation}
\begin{pmatrix}\bK_{-1} & \bP_{-1,0} & \bP_{-1,1} & \bR_{-1} \\ 
								& \bK_0 & \bP_{0,1}  & \bR_0 \\ 
								&  & \bK_1 &\bR_1 \\ 
								& & & \bK_R  \end{pmatrix} \bW
			= \omega^2 \begin{pmatrix} \bM & & \bzero& \\ & \bM & & \\ \bzero & & \bM & \\ & & & \bmm  \end{pmatrix}\bW,
\end{equation}
with the lower triangular entries in the stiffness matrix on the left being Hermitian conjugates of the corresponding upper triangular 
entries. 

The expressions for the various matrices are of the form
\begin{gather*}
K_p(m,n) = \int \left( \dfrac{d^2 N_m}{dy^2}\dfrac{d^2 N_n}{dy^2} - 2 (\kappa+ g_p)^2 \dfrac{dN_m}{dy} \dfrac{dN_n}{dy} 
							+ (\kappa + g_p)^4 N_m N_n\right) dy , \\ 
P_{p,q}(m,n) = \dfrac{k}{A_c} N_m(y_\alpha) N_n(y_\alpha) {e^{i(g_p-g_q)x_\alpha}} , \\ 
R_p(m,\alpha) = k e^{-i g_p x_\alpha} N_m(y_{\alpha}) ,\\ 
K_R = k \delta_{ij} , \;\;
M_{ij} = \delta_{ij} \rho h / D , \;\;m_{ij} = m \delta_{ij} . 
\end{gather*}
Note that $g_p$ is the $p$-th basis vector in the plane wave expansion of the plate strip. 
Traction free boundary conditions are used at the two ends of the strip for our calculations. 

\bibliographystyle{unsrt}
\bibliography{paper}

\end{document}